\documentclass[aps,prl,twocolumn,nofootinbib,balance,superscriptaddress,floats]{revtex4}

\usepackage[a4paper, total={7.2in, 10in}]{geometry}
\usepackage{latexsym}
\usepackage{dcolumn}
\usepackage{amsmath}
\usepackage{epsf}
\usepackage{float}
\usepackage{hyperref}

\usepackage[pdftex]{graphicx}
\usepackage{epstopdf}
\usepackage{pdfpages}

\usepackage{soul}

\usepackage{upgreek}
\usepackage{tabularx}

\begin{document}
\raggedbottom
%%\preprint{APS/123-QED}

\title{Nonreciprocal Inter-band Brillouin Modulation}% Force line breaks with \\

\author{Eric A. Kittlaus$^{1}$}
\noaffiliation
\author{Nils T. Otterstrom}
\affiliation{Department of Applied Physics, Yale University, New Haven, CT 06520 USA.}
\author{Prashanta Kharel}
\affiliation{Department of Applied Physics, Yale University, New Haven, CT 06520 USA.}
\author{Shai Gertler}
\affiliation{Department of Applied Physics, Yale University, New Haven, CT 06520 USA.}
\author{Peter T. Rakich$^{1}$} %,\ddagger
\noaffiliation
%\email{peter.rakich@yale.edu}
%\affiliation{Department of Applied Physics, Yale University, New Haven, CT 06520 USA.}

%\author{Eric A. Kittlaus$^{1,\dagger}$, Nils T. Otterstrom$^{1}$, and Peter T. Rakich$^{1,\ddagger}$}
%\makeatother
%\affiliation{$^1$Department of Applied Physics, Yale University, New Haven, CT 06520 USA.}

\date{\today}
%fix italics
%fix slash update figures (si sio2) and lines

%Previous integrated isolators have not been able to achieve large nonreciprocity, large operating bandwidth, and low insertion losses. This one doesn't either, but it's closer than the others.

%As a result, there is a pressing need for linear nonreciprocal devices which can be monolithically integrated into photonic circuits. 

%In contrast to prior Brillouin- and optomechanics-based schemes for nonreciprocity, the bandwidth of this scattering process is set through optical phase-matching, not acoustic or optical resonances. 

\begin{abstract}
Achieving nonreciprocal light propagation in photonic circuits is essential to control signal crosstalk and optical back-scatter. However, realizing high-fidelity nonreciprocity in low-loss integrated photonic systems remains challenging. In this paper, we experimentally demonstrate a device concept based on nonlocal acousto-optic light scattering to produce nonreciprocal single-sideband modulation and mode conversion in an integrated silicon photonic platform. In this process, a traveling-wave acoustic phonon driven via optical forces in a silicon waveguide is used to modulate light in a spatially separate waveguide through a linear inter-band Brillouin scattering process. We demonstrate up to 38 dB of nonreciprocity with 37 dB of single-sideband suppression. In contrast to prior Brillouin- and optomechanics-based schemes for nonreciprocity, the bandwidth of this scattering process is set through optical phase-matching, not acoustic or optical resonances. As a result, record-large bandwidths in excess of 125 GHz are realized, with potential for significant further improvement through optical dispersion engineering. Tunability of the nonreciprocal modulator operation wavelength over a 35 nm bandwidth is demonstrated by varying the optical pump wavelength. Such traveling-wave acousto-optic modulators provide a promising path toward the realization of broadband, low-loss isolators and circulators in integrated photonic circuits. 
\end{abstract}

%Microwave photonic filters are attractive for their flexibility and potential for integration of microwave operations onto integrated photonic chips.
%This allows for potential improvements in size, power consumption, and cost over traditional RF filters as well as direct integration into radio over fiber systems. 
%Current narrowband filters in microwave photonics are typically based on high-Q optical resonators or MEMS systems. 

\maketitle

\section{Introduction}
%The lack of mature isolator and circulator technologies in integrated optics poses a significant obstacle to the creation of complex fully-integrated photonic systems.
%ever-increasing complexity
The rapid development of complex integrated photonic circuits has led to a pressing need for robust isolator and circulator technologies to control signal routing and protect active components from back-scatter. While there have been great efforts to miniaturize existing Faraday isolators, it is fundamentally difficult to adapt these techniques to integrated systems since magneto-optic materials are intrinsically lossy and not complementary metal-oxide-semiconductor (CMOS)-compatible \protect{\cite{Sounas2017,Shoji2008,Tien11,Bi2011,Sobu13,huang16}}. Due to the large nonlinear response of integrated waveguides, approaches based on nonlinearity have been shown to permit nonreciprocal light propagation, but these are typically limited to specific operating conditions and input signals \protect{\cite{gallo01,Soljacic03,Fan2012,Shi2015}}. Other methods based on optical phase modulation have demonstrated nonreciprocity, though these techniques shift light to unwanted nearby frequency components and are only suitable for continuous-wave operation \protect{\cite{ibrahim04,Doerr11,Doerr14}}. In comparison to these approaches, an ideal nonreciprocal device will have a linear response for a large range of input powers, operate over a wide optical bandwidth, have low intrinsic optical absorption, and provide a robust and significant nonreciprocal response.

Recent approaches based on driven photonic transitions provide a promising approach toward realizing flexible on-chip nonreciprocal devices without the use of magneto-optic materials \protect{\cite{Yu2009,Kang2011,Huang11,lira12,Poulton12,Fang12,Tzuang2014,Dong2015,kim2015,Shen2016,Ruesink2016,Kim2017,Fang2017,sohn17}}. Common to these processes is spatiotemporal modulation of light in an optical waveguide or resonator that drives a transition between optical dispersion bands with a nonreciprocal phase shift or phase-matching condition. This approach to nonreciprocal modulation offers high linearity and may be realized with electro-optic, electro-mechanical, or optomechanical driving, allowing dynamic reconfigurability. As a result, such methods provide a path towards realizing flexible, broadband isolators with CMOS-compatible integration. Thus far, electro-optic implementations of inter-band photonic transitions have demonstrated large operating bandwidths for nonreciprocal light propagation, but high optical insertion losses remain a key limiter of device performance \protect{\cite{lira12}}. Opto- and electro-mechanical approaches to nonreciprocity offer the possibility for very low optical insertion losses. Thus far, however, these strategies have used resonant interactions to achieve nonreciprocal effects, which has limited operation to narrow ($\leq$1 GHz) bandwidths \protect{\cite{Kang2011,Shen2016,Ruesink2016,Kim2017,Fang2017,sohn17,shen2018reconfigurable,ruesink2018optical}. 

In this article, we demonstrate a new device that harnesses optically-driven acoustic waves to produce unidirectional optical modulation and mode conversion over nm-bandwidths. This nonreciprocal operation, realized in a low-loss integrated silicon waveguide, utilizes a nonlocal inter-band Brillouin scattering (NIBS) process in which an optically-driven traveling-wave acoustic phonon time-modulates light guided in a spatially separate optical waveguide. This process is used to produce nonreciprocal modulation with up to 38 dB of contrast between forward- and backward-propagating waves. The resulting output spectrum is single-sideband frequency-shifted with 37 dB relative suppression of spurious tones. In contrast to conventional Brillouin-based signal processing techniques, the bandwidth of this modulation process is controlled through optical phase-matching, rather than being limited by the lifetimes of resonant optical or acoustic modes; this permits operating bandwidths that are two orders of magnitude greater than state-of-the-art optomechanical modulators, and four orders of magnitude greater than the device's intrinsic acoustic response. Furthermore, by varying the wavelength of the optical pump, and by extension the driven phonon wavevector, this process can be tuned over a 35 nm bandwidth using the same device. This traveling-wave nonreciprocal modulator bridges the gap between current schemes for broadband electro-optic nonreciprocity and low-loss optomechanical modulation, representing a significant step toward the creation of broadband, high-performance integrated isolators and circulators.

\begin{figure*}[t]
\centering%\vspace{-10pt}
\includegraphics[width=\linewidth]{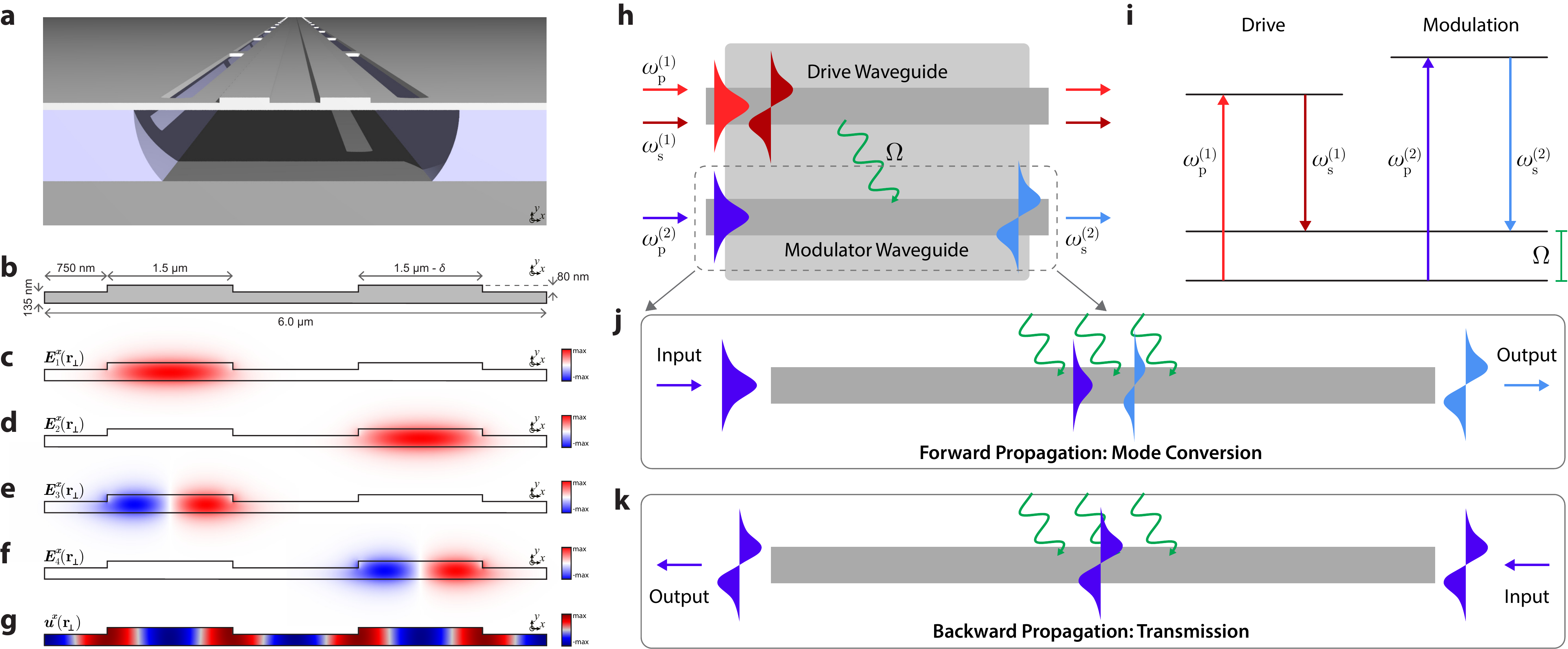}
\caption{Silicon waveguide inter-band modulator. (a) Artistic representation of device. A membrane structure with two ridge waveguides is suspended by periodic nanoscale tethers. (b) Diagram of device cross-section and dimensions. A small asymmetry $\delta$ between waveguide widths is designed to prevent optical crosstalk between the two cores. (c-f) plot the $E_x$ component of the fundamental (symmetric) and first-excited (anti-symmetric) optical modes of each waveguide, respectively. (g) plots the $x-$displacement component of the $\sim$5.7 GHz acoustic mode which mediates inter-modal Brillouin coupling in both ridge waveguides. (h) diagrams the spatial character of the nonlocal inter-band Brillouin scattering process. Two optical waves guided in different spatial modes (dispersion bands) of a `drive' waveguide transduce a monochromatic traveling-wave acoustic phonon at their difference frequency. This phonon frequency-shifts and mode-converts light guided in a spatially-separate `modulator' waveguide. (i) plots an energy level diagram for this nonlocal scattering process---note that the optical frequencies for drive and modulation processes need not be the same. (j-k) are diagrams depicting light propagation in both directions through the modulator waveguide. Forward-propagating light is mode-converted and frequency-shifted by the incident phonon as it traverses the device. By contrast, backward-propagating light propagates through the device unaffected.}
\label{fig:intro}
\end{figure*}

\section{Results}

\subsection{Silicon Waveguide Nonreciprocal Modulator}

%The mechanism of the inter-modal emit-receive process is diagrammed in Fig 1a. Strong pump waves at frequencies $\omega_{1,p}$ and $\omega_{1,s}$ are injected into distinct symmetric and anti-symmetric optical modes of an optical waveguide. These optical fields excite a travelling-wave acoustic phonon at difference frequency $\Omega = \omega_{1,p} - \omega_{1,s}$ through a stimulated inter-modal Brillouin scattering (SIMS) process. A separate optical 

We demonstrate nonreciprocal inter-band modulation utilizing the dual-core optomechanical waveguide diagrammed in Fig. \ref{fig:intro}a-b. This structure consists of a suspended silicon membrane which guides both light and sound waves. While light is confined to the cores of two distinct multi-mode ridge waveguides, guided optical waves may interact with elastic waves that extend throughout the membrane structure. Each waveguide supports a fundamental optical mode with a symmetric $E_x$-field profile (Fig. \ref{fig:intro}c-d) and a higher-order mode with an anti-symmetric field profile (Fig. \ref{fig:intro}e-f) around a vacuum wavelength $\lambda_0 \approx 1550$ nm. One acoustic phonon mode, which mediates nonlocal acousto-optic coupling around a frequency of $\Omega_\textup{B} = 5.7$ GHz, is plotted in Fig. \ref{fig:intro}g.

%These waveguides are designed to be slightly different in width by an amount $\delta.$ This asymmetry inhibits evanescent coupling between the two waveguides to ensure that they are optically isolated from one another. Each waveguide guides a fundamental mode with a symmetric $E_x$-field profile (Fig. \ref{fig:intro}c-d) and a higher-order mode with an anti-symmetric field profile (Fig. \ref{fig:data}e-f) around a vacuum wavelength $\lambda_0 \approx 1550$ nm. The entire structure also guides elastic waves due to the large impedence mismatch between silicon and air---one such acoustic phonon mode which mediates strong inter-modal Brillouin coupling around a frequency of $\Omega_\textup{B} = 5.7$ GHz is plotted in Fig. \ref{fig:intro}g.

\begin{figure*}[htbp]
\centering%\vspace{-10pt}
\includegraphics[width=.8\linewidth]{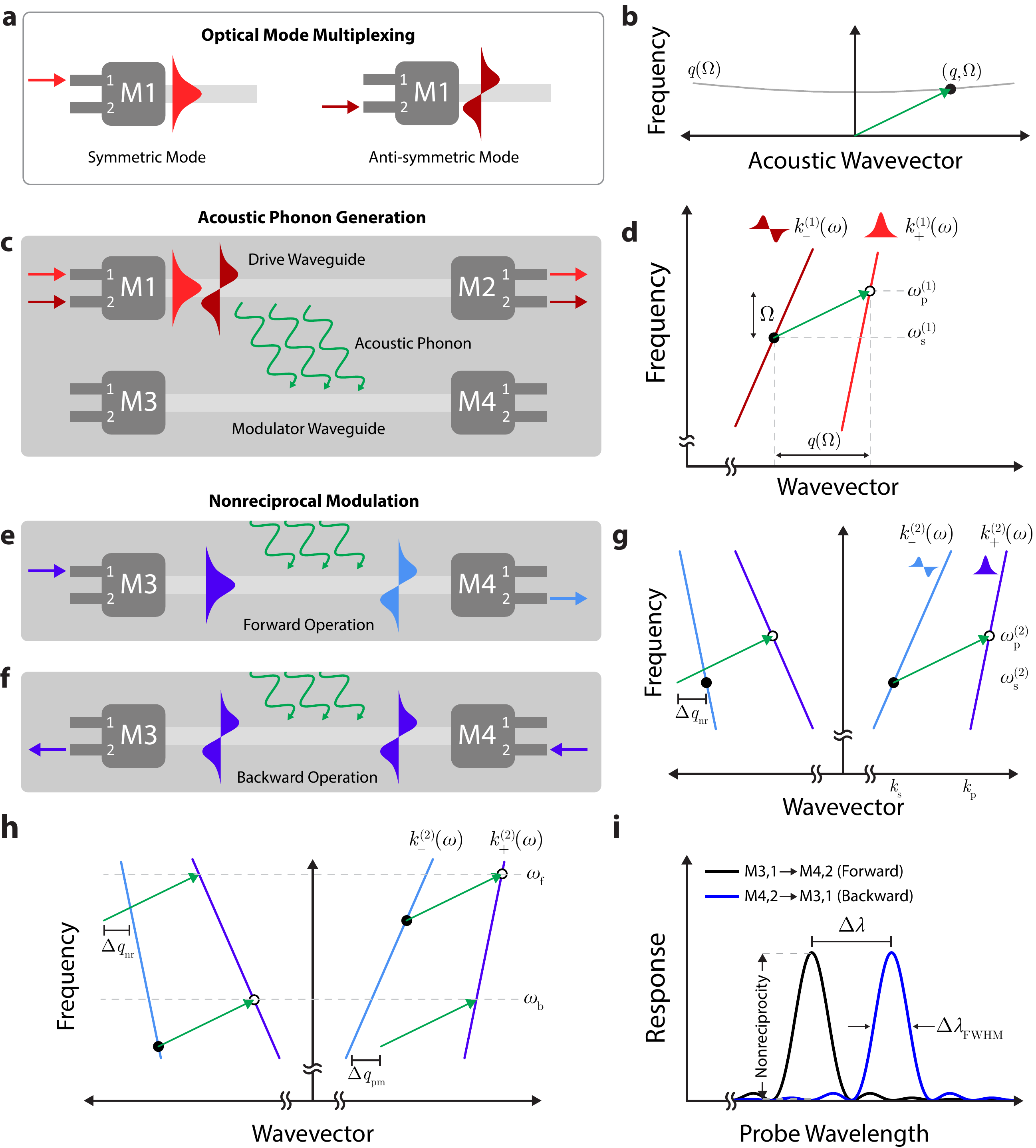}
\caption{Phase-matching and operation scheme of the inter-band Brillouin modulator. (a) diagrams the operation of mode multiplexers used to address the symmetric and anti-symmetric waveguide modes. (b) plots the dispersion relation of the acoustic phonon mode which mediates the NIBS process. (c) depicts the acoustic phonon generation process. Two strong pump waves separated by a frequencies $\omega_\textup{p}$ and $\omega_\textup{s} = \omega_\textup{p}-\Omega$ are coupled into separate optical modes of the drive waveguide. These optical fields excite a monochromatic acoustic phonon at their difference frequency $\Omega$. This phonon is then incident on the spatially distinct modulator waveguide. (d) depicts phase-matching and energy matching for SIMS. Through this process, an optically-driven acoustic phonon mediates energy transfer between initial (open circle) and final (closed circle) states on distinct optical dispersion curves through SIMS. (e-f) depict the response of the modulator waveguide when light is injected in two separate ports. In forward operation (e), light injected into port 1 of M3 is scattered from the symmetric to the anti-symmetric mode and frequency-shifted by the incident phonon before exiting the device through port 2 of M4. By contrast, in backward operation (f), light incident in port 2 of M4 is unaffected by the acoustic wave and propagates unchanged in the anti-symmetric mode before exiting the device through port 2 of M3. These behaviors can be understood through the phase-matching diagrams in (g). In the forward direction (right side), the phonon is phase-matched to a photonic transition between symmetric and anti-symmetric modes. However, in the backward direction (left side) there is a wavevector mismatch $\Delta q_\textup{nr}$ (see Eq. \ref{eq}). As a result, the same phonon does not mediate an inter-band photonic transition in the backward direction. (h) plots the effect of optical dispersion on phase-matching for these processes. Because the two optical modes do not have the same group velocity, as the wavelength of injected light is changed from the phase-matched value it accumulates a wavevector mismatch $\Delta q_\textup{pm}.$ This phase walkoff results in a finite phase-matching bandwidth for the scattering process, but can also balance the nonreciprocal wavevector mismatch $\Delta q_\textup{nr}$ to enable scattering in the backward propagation direction, resulting in the transmission spectra plotted in (i).}
\label{fig:pper}
\end{figure*}

Inter-band modulation is realized in this structure through the process diagrammed in Fig 1h. Two strong pump waves at frequencies $\omega_\textup{p}^{(1)}$ and $\omega_\textup{s}^{(1)}$ are injected into the symmetric and anti-symmetric optical modes of one of the ridge waveguides (labeled `drive' in Fig. \ref{fig:intro}h). These fields excite a monochromatic, traveling-wave acoustic phonon at difference frequency $\Omega = \omega_\textup{p}^{(1)} - \omega_\textup{s}^{(1)}$ through an inter-band Brillouin process called stimulated inter-modal Brillouin scattering (SIMS) \protect{\cite{Kittlaus2017,Otterstrom2017}}. A probe wave (frequency $\omega_\textup{p}^{(2)}$) is injected into the symmetric optical mode of a distinct `modulator' waveguide. The driven acoustic phonon spatiotemporally modulates the waveguide refractive index through the photoelastic effect to mode-convert and frequency-shift probe light to $\omega_\textup{s}^{(2)} = \omega_\textup{p}^{(2)} - \Omega.$ This process can be understood as a nonlocal form of coherent Stokes Brillouin scattering (in direct analogy to coherent Stokes Raman scattering) with the energy level diagram plotted in Fig 1i. A corresponding process of coherent anti-Stokes Brillouin scattering can also be produced (see Supplementary Note V for details). We collectively term these processes nonlocal inter-band Brillouin scattering (NIBS) to describe their salient spatial behavior and dynamics.

In the NIBS process, the travelling acoustic wave breaks the symmetry between forward- and backward-propagating optical waves to produce unidirectional mode conversion and single-sideband modulation. This process is diagrammed in Fig. \ref{fig:intro}j; when light propagates in the forward direction within the modulator waveguide, it is mode-converted and frequency-shifted through a linear acousto-optic modulation process. However, when light is injected into the same waveguide in the backward direction, it propagates through the waveguide unaffected (Fig. \ref{fig:intro}k).

\subsection{Operation Scheme}

The origin of the nonreciprocal modulation response can be understood by considering the phase-matching conditions for forward- and backward-propagating waves. In this section, we discuss phase-matching for the inter-band modulator as it relates to the experimentally predicted response of this multi-port system.

The NIBS modulator is interfaced with integrated mode multiplexers to separately address the guided optical modes of the optical ridge waveguides. A representation of the mode multiplexing process is diagrammed in Fig. \ref{fig:pper}a; light incident in port 1 of a mode multiplexer is coupled into the symmetric optical mode, whereas light incident in port 2 is coupled into the anti-symmetric optical mode. This process can be operated in reverse to de-multiplex these optical waves into single-mode bus waveguides. The drive and modulator waveguides are each interfaced with two multiplexers (labeled M1-M4 in Fig. \ref{fig:pper}c) to separately (de)multiplex these two optical modes. (For details on mode multiplexer design, see Ref. \protect{\cite{Kittlaus2017}}.)

The travelling elastic wave (group velocity $v_\textup{g,b} \sim 800$ m/s) that mediates inter-band coupling is optically-driven through the phonon generation process diagrammed in Fig. \ref{fig:pper}c. Light at frequency $\omega_\textup{p}^{(1)}$ is incident in port 1 of M1, and light at  $\omega_\textup{s}^{(1)}$ is incident in port 2. These pump waves propagate through the active region of the drive waveguide to transduce a coherent acoustic phonon with the dispersion relation depicted in Fig. \ref{fig:pper}b. In this configuration, the optical fields drive a forward-moving phonon with frequency and wavevector $(\Omega,q(\Omega)).$ This phonon satisfies both energy conservation ($\Omega = \omega_\textup{p}^{(1)} - \omega_\textup{s}^{(1)}$) and phase-matching ($q(\Omega) = k_{+}^{(1)}(\omega_\textup{p}^{(1)}) - k_{-}^{(1)}(\omega_\textup{p}^{(1)} - \Omega)$) for a SIMS process, as represented diagrammatically in Fig. \ref{fig:pper}d; here $k_{+}^{(1)}$ and $k_{-}^{(1)}$ are the dispersion relations for the symmetric and anti-symmetric optical modes in the drive waveguide, respectively. In this representation, the phonon (green arrow in Fig. \ref{fig:pper}d) scatters a photon between initial (open circle) and final (closed circle) states on separate optical dispersion bands through a SIMS process.

This driven phonon may then mediate an inter-band transition through NIBS in a spatially separate modulator waveguide, as diagrammed in Fig. \ref{fig:pper}e-g. Through this process, light at frequency $\omega_\textup{p}^{(2)}$ incident in port 1 of M3 is mode-converted and red-shifted to Stokes frequency $\omega_\textup{s}^{(2)}$ through a linear acousto-optic scattering process. After passing through the active waveguide region, this light exits the device through port 2 of M4; any residual un-shifted light remains in the symmetric mode and exits the device through port 1 of M4.  As in the drive waveguide, this Stokes scattering process must satisfy energy conservation ($\omega_\textup{p}^{(2)} - \omega_\textup{s}^{(2)} = \Omega$) and phase-matching ($ k_{+}^{(2)}(\omega_\textup{p}^{(2)}) - k_{-}^{(2)}(\omega_\textup{p}^{(2)} - \Omega) = q(\Omega)$), as shown in the right side of Fig. \ref{fig:pper}g, where $k_{+}^{(2)}$ and $k_{-}^{(2)}$ are the wavevectors of the symmetric and anti-symmetric optical modes within the modulator waveguide. Note, however, that since the drive and modulator waveguides are not necessarily identical, the optical dispersion relations and frequencies involved in this process may be very different from those used in the phonon generation process.  Within the silicon optomechanical modulator devices, where the drive and modulator waveguides differ in dimension (with a width asymmetry $\delta$ as depicted in Fig. \ref{fig:intro}b), efficient inter-modal coupling occurs when that the optical frequencies in both waveguides phase-match to scattering from the same phonon.

%the optical dispersion relations and frequencies involved in this process may be very different from those used in the phonon generation process. 
%Nonetheless, to satisfy phase-matching the wavevector difference between the two optical modes must still be equal to the phonon wavevector.  
The traveling-wave phonon breaks the symmetry between forward and backward light propagation in the modulator waveguide when it phase-matches to an inter-band transition in just one direction \protect{\cite{Yu2009,Poulton12}}. This symmetry-breaking, diagrammed in Fig. \ref{fig:pper}g, can be seen as resulting from the dispersion of the traveling optical waves. For a scattering process that is phase-matched in the forward direction, 

\begin{equation}
 k_{+}^{(2)}(\omega_\textup{p}^{(2)}) - k_{-}^{(2)}(\omega_\textup{p}^{(2)} - \Omega) = q(\Omega).
\end{equation}

However, for light is injected in the backward direction, phase-matching dictates that  

\begin{equation}
 k_{+}^{(2)}(\omega_\textup{p}^{(2)} - \Omega) - k_{-}^{(2)}(\omega_\textup{p}^{(2)}) = q(\Omega) - \Delta q_\textup{nr}.
\end{equation}

Here $\Delta q_\textup{nr}L$ is the optical phase mismatch accumulation in the backward direction after propagating through a device of length $L$. We can calculate the nonreciprocal wavevector mismatch by subtracting the phase-matching conditions for forward and backward Stokes processes to find

%= k_{+}^{(2)}(\omega_\textup{p}^{(2)}) - k_{+}^{(2)}(\omega_\textup{p}^{(2)}-\Omega) + k_{-}^{(2)}(\omega_\textup{p}^{(2)}) - k_{-}^{(2)}(\omega_\textup{p}^{(2)}-\Omega)

\begin{equation}
\Delta q_\textup{nr}  \approx \frac{\Omega}{c}\left( n_{\textup{g},+}^{(2)} + n_{\textup{g},-}^{(2)}\right),
\label{eq}
\end{equation}
where $n_{\textup{g},+}^{(2)}$ and $n_{\textup{g},-}^{(2)}$ are the optical group velocities of the two modes evaluated around $\omega_\textup{p}^{(2)}$ (for full details see Supplementary Note I.B). Provided that $\Delta q_\textup{nr} L \gg 1,$ a phonon that mediates phase-matched modulation in a given direction does not phase-match to a modulation process in the opposite direction. In this case, NIBS produces unidirectional mode conversion between the two guided modes represented by an asymmetric scattering matrix (see Supplementary Note III). 

%the inter-band scattering process will only be phase-matched in one propagation direction

\begin{figure*}[htbp]
\centering%\vspace{-10pt}
\includegraphics[width=.81\linewidth]{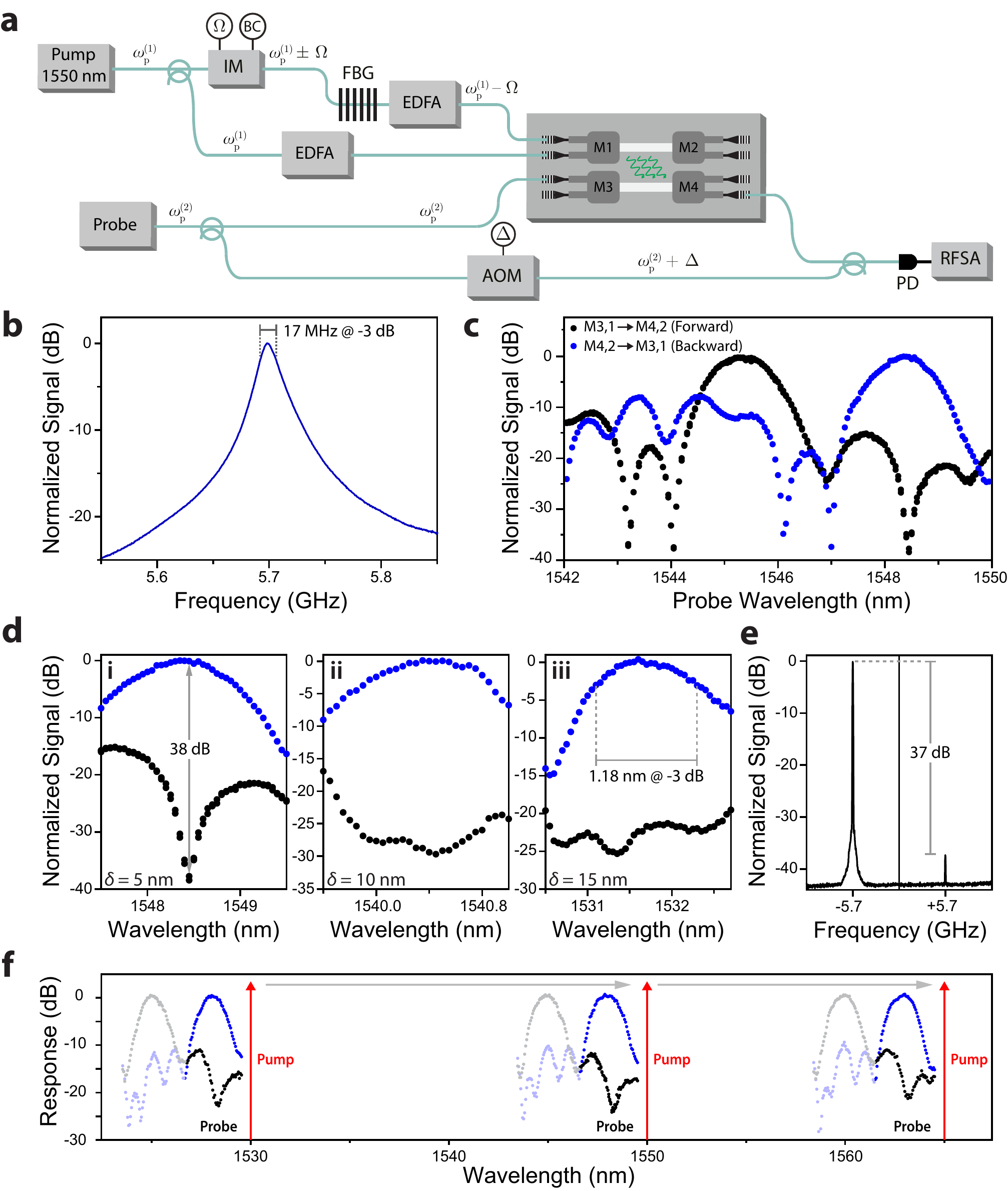}
\caption{Experimental characterization of the nonreciprocal modulator. (a) Experimental setup for measurement of the modulation response. Optical pump tones at frequencies $\omega_\textup{p}^{(1)}$ and  $\omega_\textup{s}^{(1)} = \omega_\textup{p}^{(1)} - \Omega$ are synthesized from the same laser in the upper path and incident in separate modes of the drive waveguide. A second probe laser at frequency $\omega_\textup{p}^{(2)}$ is split into two paths; in the upper arm, light is coupled into the modulator waveguide in either mode or propagation direction depending on the selected input port. After propagating through the modulator waveguide, light is coupled off-chip and combined with a frequency-shifted local oscillator (lower arm) at frequency $\omega_\textup{p}^{(2)} + \Delta$ for heterodyne spectral analysis. (b) Phase-matched frequency response of modulation for frequency-shifted probe light at $\omega_\textup{s}^{(2)} = \omega_\textup{p}^{(2)} + \Omega$ as a function of drive-wave detuning $\Omega.$ These data show strong Brillouin coupling through a resonant acoustic mode at frequency $\Omega_\textup{B}/2\pi = 5.7 $ GHz. (c) Experimental probe-wavelength dependence of the Stokes scattering efficiency when the drive-wave detuning is set to $\Omega = \Omega_\textup{B}$ for forward- and backward-injected probe light in a single device. (d) Zoomed-in plots for three different devices showing forward/backward Stokes scattering efficiency around the wavelength of optimal backward phase-matching, demonstrating nonreciprocal modulation and mode conversion. The data are for drive/modulator waveguide width asymmetries of (i) $\delta = $ 5 nm, (ii) $\delta = $ 10 nm, and (iii) $\delta = $ 15 nm. Note that as the difference in waveguide widths is increases, the difference between drive and modulation wavelengths must be increased to satisfy phase-matching. (e) plots Stokes/anti-Stokes asymmetry for scattered probe light when the modulator is driven on-resonance ($\Omega = \Omega_\textup{B}$). A small amount of output light is blue-shifted through an anti-Stokes scattering process due to crosstalk in the integrated mode multiplexers. (f) plots tuning of the probe modulation response as a function of pump laser wavelength. As the pump wavelength is tuned from 1530 nm to 1565 nm, the probe response is translated in wavelength by a corresponding amount.}
\label{fig:data}
\end{figure*}

The bandwidth over which inter-band scattering occurs is directly set by the difference in group velocities between optical modes. In comparison to fiber systems where polarization multiplexing has been explored \protect{\cite{Yu2009}}, distinct optical modes in integrated waveguides typically have significantly different optical group velocities (i.e. their dispersion bands are not parallel). As a result, as the frequency of optical probe light is changed from the center value for phase-matching, the inter-band scattering process experiences a dispersive wavevector mismatch
\begin{equation}
\Delta q_\textup{pm} \approx \frac{n_{\textup{g},-}^{(2)} - n_{\textup{g},+}^{(2)}}{c} \Delta \omega,
\end{equation}
where $\Delta \omega$ is the frequency difference between the experimental probe frequency $\omega_\textup{p}^{(2)}$ and the frequency for which phase-matching is perfectly satisfied. This results in a full-width at half-maximum operating bandwidth defined by $\Delta q_\textup{pm} L/2 = 1.39$ (for full details see Supplementary Note I.A).

Interestingly, $\Delta q_\textup{pm}$ may exactly cancel the phase mismatch between forward/backward propagation $\Delta q_\textup{nr}$, as diagrammed in Fig. \ref{fig:pper}h. This results in efficient inter-band modulation in the backward propagation direction. While phase-matching in the forward direction may be achieved at a probe frequency $\omega_\textup{p}^{(2)} = \omega_\textup{f},$ optical dispersion permits phase-matching in the backward direction at $\omega_\textup{p}^{(2)} = \omega_\textup{b}.$ The resulting two-way optical transmission spectrum between port 1 of M3 and port 2 of M4 is illustrated in Fig. \ref{fig:pper}i; each direction experiences a sinc-squared modulation response with center frequencies determined through the optical dispersion relations. Significant nonreciprocity occurs when the peak of transmission in one direction coincides with negligible transmission in the opposite propagation direction (see Supplementary Note I for full details).

\subsection{Experimental Characterization of Nonreciprocal Response}

%Pump arm: A pump laser at frequency $\omega_\textup{p}^{(1)}$ is split into two paths. In the upper path, an intensity modulator, circulator, and fiber Bragg grating are used to synthesize and select a red-shifted tone at frequency $\omega_\textup{p}^{(1)} - \Omega,$ where $\Omega$ is the RF drive frequency of the intensity modulator. This optical wave is amplified through an erbium-doped fiber amplifier and coupled into the antisymmetric mode of the emit waveguide. In the lower path, light at frequency $\omega_\textup{p}^{(1)}$ is amplified through an erbium-doped fiber amplifier before being coupled into the symmetric waveguide mode of the emit waveguide. Probe arm: Light from a probe laser is split into two paths. In the upper path, light is coupled into the receive waveguide. Depending on which mode multiplexer and port is used, probe light may be injected into either optical mode and propagation direction in this waveguide. After propagating through the receive waveguide, probe light is coupled off-chip and combined with a frequency-shifted local oscillator (lower arm) at frequency $\omega_\textup{p}^{(2)} + \Delta$ for heterodyne spectral analysis.

The silicon waveguide nonreciprocal modulator is experimentally characterized using the apparatus diagrammed in Fig. \ref{fig:data}a. Two strong pump waves (total on-chip power 90 mW) at frequencies $\omega_\textup{p}^{(1)}$ and $\omega_\textup{s}^{(1)} = \omega_\textup{p}^{(1)} - \Omega$  are synthesized from the same laser operating around 1550 nm. Light at $\omega_\textup{p}^{(1)}$ is split into two paths; one is amplified through an erbium-doped fiber amplifier and coupled into the symmetric mode of the drive waveguide. In the other path, a strong tone at $\omega_\textup{p}^{(1)} - \Omega$ is synthesized using a null-biased intensity modulator and narrowband fiber Bragg grating notch filter. The value of $\Omega$ is controlled using a microwave frequency synthesizer. This wave is amplified and coupled into the anti-symmetric mode of the drive waveguide. Probe light at frequency $\omega_\textup{p}^{(2)}$ is generated from a separate tunable laser. This light is split into two paths; in the first path, light is injected into the modulator waveguide in either the forward or backward direction. In the second probe path, light is frequency-shifted using an acousto-optic modulator to $\omega_\textup{p}^{(2)} + \Delta$ to act as an optical local oscillator. This tone is combined with the output light from the chip on a fast photodiode, where heterodyne spectral analysis is performed in the microwave domain using a radiofrequency spectrum analyzer. 

The modulation response of the device is plotted in Fig. \ref{fig:data}b-c. Fig. \ref{fig:data}b shows the frequency response of the modulated signal amplitude as a function of drive frequency $\Omega$ when $\omega_\textup{p}^{(2)}$ is set such that phase-matching is satisfied for backward-propagating light ($\Delta q = 0,$ $\lambda_\textup{p}^{(2)} = \lambda_\textup{b} = 1548.5 $ nm). These data reveal a resonant response around $\Omega/2\pi = \Omega_\textup{B}/2\pi = 5.7$ GHz corresponding to coupling mediated by the acoustic phonon mode diagrammed in Fig. \ref{fig:intro}g. For the rest of the paper, this drive frequency, and hence modulation frequency, is fixed to the acoustic phonon resonance frequency, which ensures optimal modulation efficiency. Fig. \ref{fig:data}c plots the  modulation efficiency at $\Omega = \Omega_\textup{B}$ as a function of probe wavelength $\lambda_\textup{p}^{(2)}$ for both forward and backward propagating light, showing a FWHM modulation bandwidth of about 1 nm (125 GHz) in both propagation directions. Maximum nonreciprocity is achieved between forward/backward propagation around a center wavelength of 1548.5 nm. In this configuration, significant modulation is achieved only in the backward direction. The deviation in these responses from the ideal sinc-squared response (Fig. \ref{fig:pper}i) is likely due to inhomogeneities in device fabrication (see Supplementary Note I.E for more information).  

Nonreciprocal modulation data for three different devices with the same acoustic resonance frequency are plotted in Fig. \ref{fig:data}d.i-iii. These data correspond to devices with waveguide-width asymmetries of $\delta = 5 $ nm, 10 nm, and 15 nm, respectively, with Fig. \ref{fig:data}d.i representing the same device as studied in Fig 3b-c. These data demonstrate a maximum nonreciprocity of 38 dB in Fig. \ref{fig:data}d.i, and more than 19 dB of nonreciprocity over the entire FWHM bandwidth (1.18 nm, or 150 GHz) of the device in Fig. \ref{fig:data}d.iii. Note that the center wavelength for maximum modulation is different for each device--this results from the variation in optical dispersion as device core size is changed. Using this principle, it should be possible to design modulator devices where optical drive and inter-band modulation wavelengths are very different (e.g. drive light at 1550 nm could be used to modulate a signal in the mid-infrared, or vice versa). 

In addition to varying the center modulation wavelength through device design, the wavelength response of the NIBS modulator is also directly tunable by changing the pump wavelength, and consequentially, the incident phonon wavevector. This wavelength-agility is demonstrated in Fig. \ref{fig:data}f. Within the same device, the pump wavelength is tuned from 1530 nm to 1565 nm, translating the probe modulation response by a corresponding amount with negligible changes to the overall modulation response shape. Through this process, the $\sim$1 nm operation bandwidth may be continuously tuned over the entire C band. This range was limited only by the drive laser wavelength tunability (for full details, see Supplementary Note IV).  

The nonreciprocal NIBS modulator behaves as a single-sideband frequency shifter because Stokes and anti-Stokes processes are inherently decoupled in inter-modal Brillouin scattering \protect{\cite{Kittlaus2017}}. Fig. \ref{fig:data}e plots the measured optical signal as a function of frequency relative to the incident probe wave. A Stokes/anti-Stokes asymmetry of 37 dB is demonstrated through this process. Throughout these experiments, the maximum on-chip modulation efficiency relative to the input probe power $\eta^2 \equiv P(\textup{M}3\textup{p}1)/P(\textup{M}4\textup{p}2)$ is around $1 \%$ for all tested devices (for full details see Supplementary Note VI).

\section{Discussion}

We have demonstrated an intriguing form of nonlocal Brillouin scattering and used this interaction to create nonreciprocal inter-band modulation in silicon. Through this process, we have realized high contrast (20-40 dB) non-reciprocal coupling over large operation bandwidths (150 GHz) and demonstrated the ability to tune this interaction over an unprecedented 4 THz (35 nm) frequency window. For comparison, these coupling bandwidths are approximately two orders of magnitude larger than recent demonstrations of nonreciprocal light propagation based on interactions utilizing resonant photonic modes \protect{\cite{sohn17}}. Because the operation bandwidth of NIBS is directly set by optical dispersion $(\propto \left(n_{\textup{g},-}^{(1)}-n_{\textup{g},+}^{(2)}\right)^{-1} ),$ nonreciprocal operation over bandwidths as large as 10-100 nm could be supported in traveling-wave systems through dispersion engineering  \protect{\cite{Huang11,Poulton12}}. These characteristics, combined with the low ($<$1 dB) propagation losses of our waveguide system, demonstrate the great promise and versatility of acousto-optic inter-band coupling as a basis for integrated modulator, isolator, and circulator technologies. 

While inter-band transitions have been identified as having great potential for wide-band optical isolation, experimental realizations have lagged significantly behind theoretical proposals. Early experiments sought to use electro-optic interactions to drive nonreciprocal inter-band scattering. While these studies demonstrated that coupling is possible over comparable ($\sim$200 GHz) bandwidths, technical challenges related to the patterning of complex electro-optic and metallic structures directly co-located with the optical waveguide resulted in prohibitively high (70 dB/cm) propagation losses \protect{\cite{lira12}}. 

By contrast, through use of a distributed, nonlocal acoustic emitter to drive inter-band transitions, the NIBS modulator maintains low optical propagation losses necessary for robust isolator and circulator technologies. Furthermore, this distributed drive provides the fidelity (i.e. uniform wavevector emission, travelling-wave phase-matching, etc.) and control necessary to yield high contrast nonreciprocity. This device supports estimated linear propagation losses of 0.2 dB/cm and 0.4 dB/cm for symmetric and anti-symmetric optical modes, respectively, corresponding to total insertion losses less than 1 dB for this $L = 2.39$ cm long device. Moreover, by utilizing optical driving of acoustic waves, we are able to directly tune the wavevector of the phonons which mediate inter-band modulation, allowing us to tune the wavelength of nonreciprocal device operation across the C band. 

To utilize such inter-band coupling processes as a basis for low-loss isolators and circulators, it is necessary to push inter-band scattering efficiencies close to unity. Through these experiments, we have demonstrated an inter-band coupling (mode conversion) efficiency of $\eta^2 \approx 1\%$ using a a total guided-wave power $P \approx 90$ mW within the drive waveguide. To achieve unity efficiency with optical driving of acoustic phonons, new designs to increase phonon generation efficiency or improve optical power handling may be necessary. For example, within the current device framework, optical power handling (and hence phonon drive intensity) could be greatly improved through free-carrier extraction \protect{\cite{rong2005}, or by using pump wavelengths above $\lambda = 2.1 $ $\upmu$m, where two-photon absorption vanishes in silicon.

Electromechanical phonon transduction provides an alternate approach for efficient inter-band coupling. While such schemes based on piezoelectricity are not typically wavevector-tunable (and hence fixed in operating wavelength), they permit efficient acoustic transduction. In the current silicon NIBS modulator, optically-driven phonon powers $P_\textup{b} \approx 0.1$ nW are sufficient to permit mode conversion efficiency of up to $1\%,$ owning to silicon's large acousto-optic figure of merit. Because this conversion efficiency scales linearly with acoustic power \protect{\cite{Tadesse2014,sohn17,yariv1975quantum}}, piezoelectric transduction, which is routinely used to produce acoustic powers $\gg$1 mW \protect{\cite{kuhn71,sasaki74,Ohmachi77}}, provides a natural path towards unity-efficiency inter-band coupling. Hence, innovate electromechanical transducer designs that mimic the distributed phonon emission produced within the optomechanical NIBS modulator could offer a compelling path toward broadband, low-loss, and energy-efficient acousto-optic isolators.

In summary, we have demonstrated nonreciprocal single-sideband optical modulation in a silicon waveguide through a nonlocal inter-band Brillouin scattering process. This device produces record-wide bandwidths for inter-band optomechanical modulation while supporting significant (20-40 dB) nonreciprocal contrast in a low-loss silicon waveguide. In contrast to prior schemes for acousto-optic nonreciprocity, the operation bandwidth of this process is set by optical dispersion, and not limited by the lifetimes of resonant phononic or photonic modes. As a result, this bandwidth can be extended by orders of magnitude through optical dispersion engineering \protect{\cite{Huang11,Poulton12}. This approach may enable the creation of ultra-broadband, low-loss nonreciprocal modulators, circulators, and isolators in silicon photonic circuits.

\section{Methods}

\subsection{Device Fabrication}
The suspended optomechanical waveguide structures were written through a two-step lithography process. First, ridge waveguides were patterned on a silicon-on-insulator chip with a 215 nm crystalline silicon top layer using electron beam lithography on hydrogen silsesquioxane photoresist. After development, a Cl$_2$ reactive ion etch (RIE) was employed to etch the ridge waveguides and grating couplers. In the second lithography step, slots were written to expose the oxide layer using electron beam lithography of ZEP520A photoresist and Cl$_2$ RIE. The oxide undercladding was then removed through a wet etch in 49\% hydrofluoric acid. The device under test is comprised of 468 suspended segments 50 $\upmu$m in length.

\subsection{Experiment}
Light is coupled on- and off-chip using commercially-manufactured four-port fiber arrays and integrated grating couplers, with fiber-to-chip coupling losses of 7 dB/facet. The following abbreviations are used in the experimental diagrams: IM Mach-Zehnder intensity modulator, BC bias controller, EDFA erbium-doped fiber amplifier, AOM acousto-optic frequency shifter, FBG fiber Bragg grating, PD photodetector, RFSA radio-frequency spectrum analyzer.  

\subsection{Acknowledgements}
This work was supported through a seedling grant under the direction of Dr. Daniel Green at DARPA MTO and by the Packard Fellowship for Science and Engineering; N.T.O. acknowledges support from the National Science Foundation Graduate Research Fellowship under Grant No. DGE1122492. 

%\subsection{Author Contributions}
%E.A.K. designed and fabricated the waveguide devices. E.A.K., P.K., S.G., N.T.O. and P.T.R. developed numerical and analytical models of the device physics. E.A.K., N.T.O., and S.G. conducted experiments with the assistance of P.K. and P.T.R. All authors contributed to the writing of this paper.

%\subsubsection{Additional Information}
%\noindent \textbf{Competing financial interests:} The authors declare no competing financial interests.

%\noindent \textbf{Data availability:} All data supporting the findings of this work are available within the article and its supplementary information files.

%\bibliographystyle{naturemag-ed} 
%\bibliography{cites}

\begin{thebibliography}{10}
\expandafter\ifx\csname url\endcsname\relax
  \def\url#1{\texttt{#1}}\fi
\expandafter\ifx\csname urlprefix\endcsname\relax\def\urlprefix{URL }\fi
\providecommand{\bibinfo}[2]{#2}
\providecommand{\eprint}[2][]{\url{#2}}

\bibitem{Sounas2017}
\bibinfo{author}{Sounas, D.L.} \& \bibinfo{author}{Al{\`u}, A.}
\newblock \bibinfo{title}{Non-reciprocal photonics based on time modulation}.
\newblock \emph{\bibinfo{journal}{Nat. Photonics}}
  \textbf{\bibinfo{volume}{11}}, \bibinfo{pages}{774--783}
  (\bibinfo{year}{2017}).

\bibitem{Shoji2008}
\bibinfo{author}{Shoji, Y.}, \bibinfo{author}{Mizumoto, T.},
  \bibinfo{author}{Yokoi, H.}, \bibinfo{author}{Hsieh, I.W.} \&
  \bibinfo{author}{{Osgood Jr.}, R.M.}
\newblock \bibinfo{title}{Magneto-optical isolator with silicon waveguides
  fabricated by direct bonding}.
\newblock \emph{\bibinfo{journal}{Appl. Phys. Lett.}}
  \textbf{\bibinfo{volume}{92}}, \bibinfo{pages}{071117}
  (\bibinfo{year}{2008}).

\bibitem{Tien11}
\bibinfo{author}{Tien, M.C.}, \bibinfo{author}{Mizumoto, T.},
  \bibinfo{author}{Pintus, P.}, \bibinfo{author}{Kromer, H.} \&
  \bibinfo{author}{Bowers, J.E.}
\newblock \bibinfo{title}{Silicon ring isolators with bonded nonreciprocal
  magneto-optic garnets}.
\newblock \emph{\bibinfo{journal}{Opt. Express}} \textbf{\bibinfo{volume}{19}},
  \bibinfo{pages}{11740} (\bibinfo{year}{2011}).

\bibitem{Bi2011}
\bibinfo{author}{Bi, L.} \emph{et~al.}
\newblock \bibinfo{title}{On-chip optical isolation in monolithically
  integrated non-reciprocal optical resonators}.
\newblock \emph{\bibinfo{journal}{Nat. Photonics}}
  \textbf{\bibinfo{volume}{5}}, \bibinfo{pages}{758} (\bibinfo{year}{2011}).

\bibitem{Sobu13}
\bibinfo{author}{Sobu, Y.}, \bibinfo{author}{Shoji, Y.},
  \bibinfo{author}{Sakurai, K.} \& \bibinfo{author}{Mizumoto, T.}
\newblock \bibinfo{title}{Ga{I}n{A}s{P}/{I}n{P} {M}{Z}{I} waveguide optical
  isolator integrated with spot size converter}.
\newblock \emph{\bibinfo{journal}{Opt. Express}} \textbf{\bibinfo{volume}{21}},
  \bibinfo{pages}{15373} (\bibinfo{year}{2013}).

\bibitem{huang16}
\bibinfo{author}{Huang, D.} \emph{et~al.}
\newblock \bibinfo{title}{Electrically driven and thermally tunable integrated
  optical isolators for silicon photonics}.
\newblock \emph{\bibinfo{journal}{IEEE J. Sel. Top. Quantum Electron}}
  \textbf{\bibinfo{volume}{22}}, \bibinfo{pages}{271--278}
  (\bibinfo{year}{2016}).

\bibitem{gallo01}
\bibinfo{author}{Gallo, K.}, \bibinfo{author}{Assanto, G.},
  \bibinfo{author}{Parameswaran, K.R.} \& \bibinfo{author}{Fejer, M.M.}
\newblock \bibinfo{title}{All-optical diode in a periodically poled lithium
  niobate waveguide}.
\newblock \emph{\bibinfo{journal}{Appl. Phys. Lett.}}
  \textbf{\bibinfo{volume}{79}}, \bibinfo{pages}{314--316}
  (\bibinfo{year}{2001}).

\bibitem{Soljacic03}
\bibinfo{author}{Solja\v{c}i\'{c}, M.}, \bibinfo{author}{Luo, C.},
  \bibinfo{author}{Joannopoulos, J.D.} \& \bibinfo{author}{Fan, S.}
\newblock \bibinfo{title}{Nonlinear photonic crystal microdevices for optical
  integration}.
\newblock \emph{\bibinfo{journal}{Opt. Lett.}} \textbf{\bibinfo{volume}{28}},
  \bibinfo{pages}{637--639} (\bibinfo{year}{2003}).

\bibitem{Fan2012}
\bibinfo{author}{Fan, L.} \emph{et~al.}
\newblock \bibinfo{title}{An all-silicon passive optical diode}.
\newblock \emph{\bibinfo{journal}{Science}} \textbf{\bibinfo{volume}{335}},
  \bibinfo{pages}{447--450} (\bibinfo{year}{2012}).

\bibitem{Shi2015}
\bibinfo{author}{Shi, Y.}, \bibinfo{author}{Yu, Z.} \& \bibinfo{author}{Fan,
  S.}
\newblock \bibinfo{title}{Limitations of nonlinear optical isolators due to
  dynamic reciprocity}.
\newblock \emph{\bibinfo{journal}{Nat. Photonics}}
  \textbf{\bibinfo{volume}{9}}, \bibinfo{pages}{388} (\bibinfo{year}{2015}).

\bibitem{ibrahim04}
\bibinfo{author}{Ibrahim, S.K.}, \bibinfo{author}{Bhandare, S.},
  \bibinfo{author}{Sandel, D.}, \bibinfo{author}{Zhang, H.} \&
  \bibinfo{author}{Noe, R.}
\newblock \bibinfo{title}{Non-magnetic 30 d{B} integrated optical isolator in
  {I}{I}{I}/{V} material}.
\newblock \emph{\bibinfo{journal}{Electron. Lett.}}
  \textbf{\bibinfo{volume}{40}}, \bibinfo{pages}{1293--1294}
  (\bibinfo{year}{2004}).

\bibitem{Doerr11}
\bibinfo{author}{Doerr, C.R.}, \bibinfo{author}{Dupuis, N.} \&
  \bibinfo{author}{Zhang, L.}
\newblock \bibinfo{title}{Optical isolator using two tandem phase modulators}.
\newblock \emph{\bibinfo{journal}{Opt. Lett.}} \textbf{\bibinfo{volume}{36}},
  \bibinfo{pages}{4293--4295} (\bibinfo{year}{2011}).

\bibitem{Doerr14}
\bibinfo{author}{Doerr, C.R.}, \bibinfo{author}{Chen, L.} \&
  \bibinfo{author}{Vermeulen, D.}
\newblock \bibinfo{title}{Silicon photonics broadband modulation-based
  isolator}.
\newblock \emph{\bibinfo{journal}{Opt. Express}} \textbf{\bibinfo{volume}{22}},
  \bibinfo{pages}{4493--4498} (\bibinfo{year}{2014}).

\bibitem{Yu2009}
\bibinfo{author}{Yu, Z.} \& \bibinfo{author}{Fan, S.}
\newblock \bibinfo{title}{Complete optical isolation created by indirect
  interband photonic transitions}.
\newblock \emph{\bibinfo{journal}{Nat. Photonics}}
  \textbf{\bibinfo{volume}{3}}, \bibinfo{pages}{91--94} (\bibinfo{year}{2009}).

\bibitem{Kang2011}
\bibinfo{author}{Kang, M.S.}, \bibinfo{author}{Butsch, A.} \&
  \bibinfo{author}{Russell, P.{\relax St}.J.}
\newblock \bibinfo{title}{Reconfigurable light-driven opto-acoustic isolators
  in photonic crystal fibre}.
\newblock \emph{\bibinfo{journal}{Nat. Photonics}}
  \textbf{\bibinfo{volume}{5}}, \bibinfo{pages}{549--553}
  (\bibinfo{year}{2011}).

\bibitem{Huang11}
\bibinfo{author}{Huang, X.} \& \bibinfo{author}{Fan, S.}
\newblock \bibinfo{title}{Complete all-optical silica fiber isolator via
  stimulated {B}rillouin scattering}.
\newblock \emph{\bibinfo{journal}{J. Lightwave Technol.}}
  \textbf{\bibinfo{volume}{29}}, \bibinfo{pages}{2267--2275}
  (\bibinfo{year}{2011}).

\bibitem{lira12}
\bibinfo{author}{Lira, H.}, \bibinfo{author}{Yu, Z.}, \bibinfo{author}{Fan, S.}
  \& \bibinfo{author}{Lipson, M.}
\newblock \bibinfo{title}{Electrically driven nonreciprocity induced by
  interband photonic transition on a silicon chip}.
\newblock \emph{\bibinfo{journal}{Phys. Rev. Lett.}}
  \textbf{\bibinfo{volume}{109}}, \bibinfo{pages}{033901}
  (\bibinfo{year}{2012}).

\bibitem{Poulton12}
\bibinfo{author}{Poulton, C.G.} \emph{et~al.}
\newblock \bibinfo{title}{Design for broadband on-chip isolator using
  stimulated {B}rillouin scattering in dispersion-engineered chalcogenide
  waveguides}.
\newblock \emph{\bibinfo{journal}{Opt. Express}} \textbf{\bibinfo{volume}{20}},
  \bibinfo{pages}{21235--21246} (\bibinfo{year}{2012}).

\bibitem{Fang12}
\bibinfo{author}{Fang, K.}, \bibinfo{author}{Yu, Z.} \& \bibinfo{author}{Fan,
  S.}
\newblock \bibinfo{title}{Photonic {A}haronov-{B}ohm effect based on dynamic
  modulation}.
\newblock \emph{\bibinfo{journal}{Phys. Rev. Lett.}}
  \textbf{\bibinfo{volume}{108}}, \bibinfo{pages}{153901}
  (\bibinfo{year}{2012}).

\bibitem{Tzuang2014}
\bibinfo{author}{Tzuang, L.D.}, \bibinfo{author}{Fang, K.},
  \bibinfo{author}{Nussenzveig, P.}, \bibinfo{author}{Fan, S.} \&
  \bibinfo{author}{Lipson, M.}
\newblock \bibinfo{title}{Non-reciprocal phase shift induced by an effective
  magnetic flux for light}.
\newblock \emph{\bibinfo{journal}{Nat. Photonics}}
  \textbf{\bibinfo{volume}{8}}, \bibinfo{pages}{701} (\bibinfo{year}{2014}).

\bibitem{Dong2015}
\bibinfo{author}{Dong, C.} \emph{et~al.}
\newblock \bibinfo{title}{Brillouin-scattering-induced transparency and
  non-reciprocal light storage}.
\newblock \emph{\bibinfo{journal}{Nat. Commun.}} \textbf{\bibinfo{volume}{6}},
  \bibinfo{pages}{6193} (\bibinfo{year}{2015}).

\bibitem{kim2015}
\bibinfo{author}{Kim, J.}, \bibinfo{author}{Kuzyk, M.C.}, \bibinfo{author}{Han,
  K.}, \bibinfo{author}{Wang, H.} \& \bibinfo{author}{Bahl, G.}
\newblock \bibinfo{title}{Non-reciprocal {B}rillouin scattering induced
  transparency}.
\newblock \emph{\bibinfo{journal}{Nat. Phys.}} \textbf{\bibinfo{volume}{11}},
  \bibinfo{pages}{275--280} (\bibinfo{year}{2015}).

\bibitem{Shen2016}
\bibinfo{author}{Shen, Z.} \emph{et~al.}
\newblock \bibinfo{title}{Experimental realization of optomechanically induced
  non-reciprocity}.
\newblock \emph{\bibinfo{journal}{Nat. Photonics}}
  \textbf{\bibinfo{volume}{10}}, \bibinfo{pages}{657} (\bibinfo{year}{2016}).

\bibitem{Ruesink2016}
\bibinfo{author}{Ruesink, F.}, \bibinfo{author}{Miri, M.A.},
  \bibinfo{author}{Al{\`u}, A.} \& \bibinfo{author}{Verhagen, E.}
\newblock \bibinfo{title}{Nonreciprocity and magnetic-free isolation based on
  optomechanical interactions}.
\newblock \emph{\bibinfo{journal}{Nat. Commun.}} \textbf{\bibinfo{volume}{7}},
  \bibinfo{pages}{13662} (\bibinfo{year}{2016}).

\bibitem{Kim2017}
\bibinfo{author}{Kim, J.}, \bibinfo{author}{Kim, S.} \& \bibinfo{author}{Bahl,
  G.}
\newblock \bibinfo{title}{Complete linear optical isolation at the microscale
  with ultralow loss}.
\newblock \emph{\bibinfo{journal}{Sci. Rep.}} \textbf{\bibinfo{volume}{7}},
  \bibinfo{pages}{1647} (\bibinfo{year}{2017}).

\bibitem{Fang2017}
\bibinfo{author}{Fang, K.} \emph{et~al.}
\newblock \bibinfo{title}{Generalized non-reciprocity in an optomechanical
  circuit via synthetic magnetism and reservoir engineering}.
\newblock \emph{\bibinfo{journal}{Nat. Phys.}} \textbf{\bibinfo{volume}{13}},
  \bibinfo{pages}{465} (\bibinfo{year}{2017}).

\bibitem{sohn17}
\bibinfo{author}{Sohn, D.B.}, \bibinfo{author}{Kim, S.} \&
  \bibinfo{author}{Bahl, G.}
\newblock \bibinfo{title}{Time-reversal symmetry breaking with acoustic pumping
  of nanophotonic circuits}.
\newblock \emph{\bibinfo{journal}{Nat. Photonics}}
  \textbf{\bibinfo{volume}{12}}, \bibinfo{pages}{91} (\bibinfo{year}{2018}).

\bibitem{shen2018reconfigurable}
\bibinfo{author}{Shen, Z.} \emph{et~al.}
\newblock \bibinfo{title}{Reconfigurable optomechanical circulator and
  directional amplifier}.
\newblock \emph{\bibinfo{journal}{Nat. Commun.}} \textbf{\bibinfo{volume}{9}},
  \bibinfo{pages}{1797} (\bibinfo{year}{2018}).

\bibitem{ruesink2018optical}
\bibinfo{author}{Ruesink, F.}, \bibinfo{author}{Mathew, J.P.},
  \bibinfo{author}{Miri, M.A.}, \bibinfo{author}{Al{\`u}, A.} \&
  \bibinfo{author}{Verhagen, E.}
\newblock \bibinfo{title}{Optical circulation in a multimode optomechanical
  resonator}.
\newblock \emph{\bibinfo{journal}{Nat. Commun.}} \textbf{\bibinfo{volume}{9}},
  \bibinfo{pages}{1798} (\bibinfo{year}{2018}).

\bibitem{Kittlaus2017}
\bibinfo{author}{Kittlaus, E.A.}, \bibinfo{author}{Otterstrom, N.T.} \&
  \bibinfo{author}{Rakich, P.T.}
\newblock \bibinfo{title}{On-chip inter-modal {B}rillouin scattering}.
\newblock \emph{\bibinfo{journal}{Nat. Commun.}} \textbf{\bibinfo{volume}{8}},
  \bibinfo{pages}{15819} (\bibinfo{year}{2017}).

\bibitem{Otterstrom2017}
\bibinfo{author}{Otterstrom, N.T.}, \bibinfo{author}{Behunin, R.O.},
  \bibinfo{author}{Kittlaus, E.A.}, \bibinfo{author}{Wang, Z.} \&
  \bibinfo{author}{Rakich, P.T.}
\newblock \bibinfo{title}{A silicon {B}rillouin laser}.
\newblock \emph{\bibinfo{journal}{Science}} \textbf{\bibinfo{volume}{360}},
  \bibinfo{pages}{1113--1116} (\bibinfo{year}{2018}).

\bibitem{rong2005}
\bibinfo{author}{Rong, H.} \emph{et~al.}
\newblock \bibinfo{title}{A continuous-wave {R}aman silicon laser}.
\newblock \emph{\bibinfo{journal}{Nature}} \textbf{\bibinfo{volume}{433}},
  \bibinfo{pages}{725--728} (\bibinfo{year}{2005}).

\bibitem{Tadesse2014}
\bibinfo{author}{Tadesse, S.A.} \& \bibinfo{author}{Li, M.}
\newblock \bibinfo{title}{Sub-optical wavelength acoustic wave modulation of
  integrated photonic resonators at microwave frequencies}.
\newblock \emph{\bibinfo{journal}{Nat. Commun.}} \textbf{\bibinfo{volume}{5}},
  \bibinfo{pages}{5402} (\bibinfo{year}{2014}).

\bibitem{yariv1975quantum}
\bibinfo{author}{Yariv, A.}
\newblock \emph{\bibinfo{title}{Quantum Electronics}} (\bibinfo{publisher}{John
  Wiley \& Sons, Incorporated}, \bibinfo{year}{1989}).

\bibitem{kuhn71}
\bibinfo{author}{Kuhn, L.}, \bibinfo{author}{Heidrich, P.F.} \&
  \bibinfo{author}{Lean, E.G.}
\newblock \bibinfo{title}{Optical guided wave mode conversion by an acoustic
  surface wave}.
\newblock \emph{\bibinfo{journal}{Appl. Phys. Lett.}}
  \textbf{\bibinfo{volume}{19}}, \bibinfo{pages}{428} (\bibinfo{year}{1971}).

\bibitem{sasaki74}
\bibinfo{author}{Sasaki, H.}, \bibinfo{author}{Kushibiki, J.} \&
  \bibinfo{author}{Chubachi, N.}
\newblock \bibinfo{title}{Efficient acoustoâ€optic
  {T}{E}$\rightleftharpoons${T}{M} mode conversion in {Z}n{O} films}.
\newblock \emph{\bibinfo{journal}{Appl. Phys. Lett.}}
  \textbf{\bibinfo{volume}{25}}, \bibinfo{pages}{476--477}
  (\bibinfo{year}{1974}).

\bibitem{Ohmachi77}
\bibinfo{author}{Ohmachi, Y.} \& \bibinfo{author}{Noda, J.}
\newblock \bibinfo{title}{Li{N}b{O}3 {T}{E}{-}{T}{M} mode converter using
  collinear acoustooptic interaction}.
\newblock \emph{\bibinfo{journal}{IEEE J. Quantum Electron.}}
  \textbf{\bibinfo{volume}{13}}, \bibinfo{pages}{43--46}
  (\bibinfo{year}{1977}).

\end{thebibliography}

\newpage
\clearpage



\section*{Supplementary material}

# Supplementary Information: Nonreciprocal Inter-band Brillouin Modulation

Eric A. Kittlaus[1], Nils T. Otterstrom,[1] Prashanta Kharel,[1] Shai Gertler,[1] and Peter T. Rakich[1]

[1]*Department of Applied Physics, Yale University, New Haven, CT 06520 USA.*

(Dated: April 27, 2018)

## Contents

# I. PHASE-MATCHING AND NONRECIPROCITY

The nonreciprocal behavior of the inter-band modulation process is determined by the dispersion of the participating optical waves [1–4]. Here we explore the various relationships between device optical parameters, device operation bandwidth, and necessary conditions for significant nonreciprocal behavior.

## A. Inter-band Modulator Phase-Matching Bandwidth

We first explore the bandwidth of device operation for the inter-band photonic modulator. In particular, here we derive the bandwidth over which an incident phonon can scatter and frequency-shift light via an inter-band Brillouin scattering process.

We consider an incident acoustic phonon with frequency $\Omega$ which is perfectly phase-matched to a Stokes scattering process between two optical dispersion branches, $k_+(\omega)$ and $k_-(\omega-\Omega)$, at an optical probe frequency of $\omega = \omega_\mathrm{p}$. This process is diagrammed in Fig. 1b. In this case, the phase-matching condition reads

$$q(\Omega) = k_+(\omega_\mathrm{p}) - k_-(\omega_\mathrm{p} - \Omega) \tag{1}$$

where $q(\Omega)$ is the dispersion relation of the acoustic phonon which mediates this process (Fig. 1a). This phase-matching condition requires that the sum of the wavevectors of the initial particle states is equal to the sum of the wavevectors of the final particle states.

We simplify notation by writing the frequency-dependent difference between pump and Stokes optical wavevectors as

$$\Delta k(\omega, \Omega) = k_+(\omega) - k_-(\omega - \Omega) \tag{2}$$

so that the phase-matching condition can be rewritten as:

$$\Delta k(\omega_\mathrm{p}, \Omega) - q(\Omega) = 0 \tag{3}$$

Due to optical dispersion, as the probe frequency is detuned from $\omega_\mathrm{p}$, this equation is no longer satisfied (right side of Fig. 1c), instead resulting in a wavevector mismatch for the scattering process

$$\Delta k(\omega, \Omega) - q(\Omega) = \Delta q_\mathrm{pm}. \tag{4}$$

As light propagates through the active device region over a length $L$, this results in an accumulated phase mismatch $\Delta q_\mathrm{pm} L$ for the inter-band scattering process. We can write the frequency-dependent wavevector mismatch relative to $\omega_\mathrm{p}$ as

$$\Delta q_{\mathrm{pm}} = \Delta k(\omega, \Omega) - \Delta k(\omega_{\mathrm{p}}, \Omega) = (k_+(\omega) - k_+(\omega_{\mathrm{p}})) - (k_-(\omega - \Omega) - k_-(\omega_{\mathrm{p}} - \Omega)). \tag{5}$$

Assuming linear dispersion (i.e. constant optical group velocity) over the entire phase-matching bandwidth (an excellent approximation for typical integrated systems), we can Taylor expand to first order around $\omega$ to find:

$$\Delta q_{\mathrm{pm}} = \frac{\partial k_+}{\partial \omega}(\omega - \omega_{\mathrm{p}}) - \frac{\partial k_-}{\partial \omega}(\omega - \omega_{\mathrm{p}}) \tag{6}$$

or

$$\Delta q_{\mathrm{pm}} = \frac{n_{\mathrm{g},+} - n_{\mathrm{g},-}}{c}\Delta\omega. \tag{7}$$

Here we have defined $n_{\mathrm{g},+}$ and $n_{\mathrm{g},-}$ to be the group velocities of the two optical modes, and $\Delta\omega = \omega - \omega_{\mathrm{p}}$ to be the frequency difference between the experimental probe frequency and the frequency for which light is perfectly phase-matched to a scattering process. Note that the frequency-dependent phase mismatch is minimized when the optical group velocities of the two modes are equal (i.e. when their dispersion curves are parallel at the operating frequency).

For a device of finite length, the resulting modulation strength has a sinc-squared response $\propto \mathrm{sinc}^2(\Delta q_{\mathrm{pm}} L/2)$ (see Section 1E). This response envelope is equal to $1/2$ when $\Delta q L/2 = 1.39$ and has nulls at $\Delta q L/2 = n\pi$, where $n$ is an integer.

Therefore, the full-width at half-maximum of the modulation response is

$$\Delta\omega_{\mathrm{FWHM}} = 2\Delta\omega = \frac{4 \cdot 1.39 c}{L} \frac{1}{|n_{\mathrm{g},+} - n_{\mathrm{g},-}|}, \tag{8}$$

which can be written in units of frequency as:

$$\Delta f_{\mathrm{FWHM}} = 2\Delta f = \frac{2 \cdot 1.39 c}{\pi L} \frac{1}{|n_{\mathrm{g},+} - n_{\mathrm{g},-}|}. \tag{9}$$

This quantity represents the operating bandwidth of the inter-band photonic modulator.

### B. Forward/Backward Modulation Phase Mismatch

We next derive the modulation wavevector (and hence phase) mismatch between scattering processes for light propagating in the forward and backward directions of the inter-band modulator. This direction-dependent phase mismatch permits the nonreciprocal response of the NIBS process.

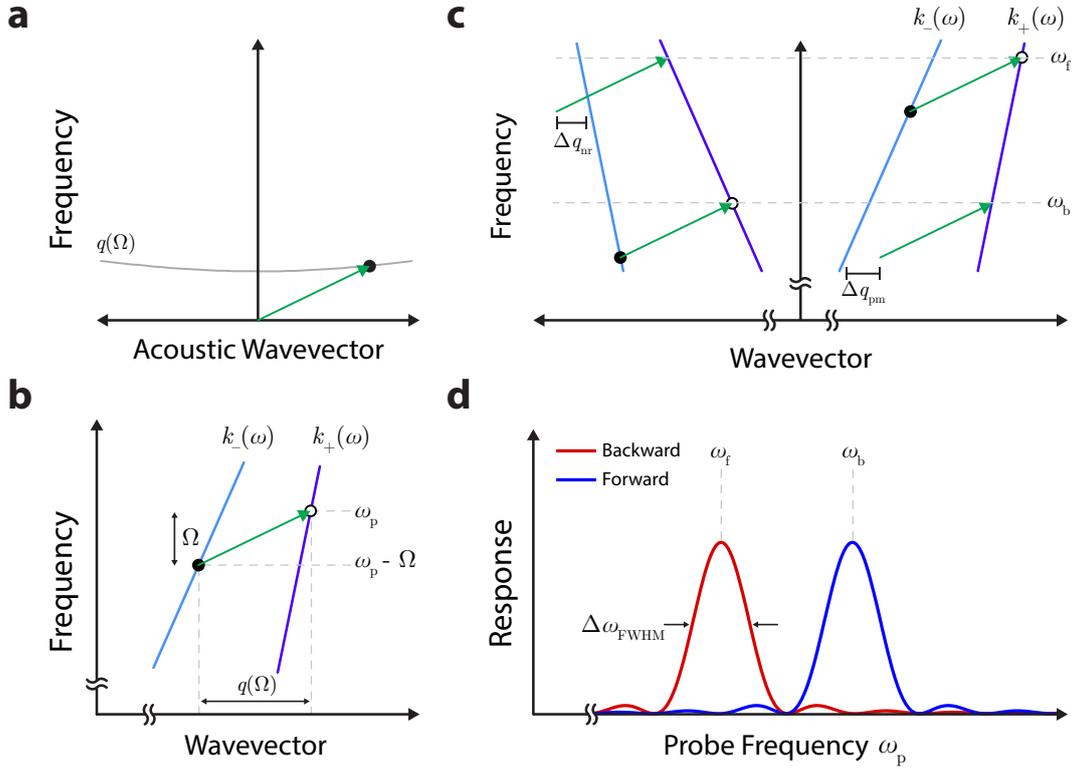

FIG. 1: Phase matching for the NIBS process. (a) Depicts the dispersion relation for the acoustic phonon which mediates the inter-band photonic transition. (b) Depicts phase-matching for the inter-band scattering process. An acoustic phonon with frequency and wavevector $(\Omega, q(\Omega))$ mediates coupling between points on two distinct optical dispersion bands at $k_+(\omega_p)$ and $k_-(\omega_p - \Omega)$ (c) Shows how this process is perfectly phase-matched only at a single frequency $\omega = \omega_f$ when these two dispersion bands have different group velocities. This results in a frequency-dependent wavevector mismatch $\Delta q_{pm}$ as $\omega$ is detuned from $\omega_f$. Furthermore, for the same incident phonon, light injected in the backward direction (left side of plot) at $\omega_f$ is not phase-matched to an inter-band transition, due to a wavevector mismatch $\Delta q_{nr}$. This nonreciprocal response results because the traveling-wave phonon breaks the symmetry between forward- and backward-propagating optical waves. However, the two sources of phase mismatch, $\Delta q_{pm}$ and $\Delta q_{nr}$ can exactly cancel, resulting in a phase-matched photonic transition in the backward direction at a frequency $\omega_b$. (d) Plots the expected modulation response (inter-band power conversion efficiency) of the device resulting from the interplay of these effects when light is injected in either the forward or backward direction.

Through operation of the photonic modulator, light injected into the device in the forward direction is mode-converted via an incident phonon. By contrast, light injected in the backward direction at the same frequency is not affected by this phonon if the inter-band scattering process is not phase-matched. We discuss this situation (diagrammed in Fig. 1c) here.

We again consider light propagating in the forward direction with a frequency $\omega_f$ that is phase-matched to a scattering process through an incident phonon with frequency $\Omega$, as in Fig. 2h of the main text. The phase-matching condition reads:

$$q(\Omega) = k_+(\omega_{\text{f}}) - k_-(\omega_{\text{f}} - \Omega) \tag{10}$$

However, for light injected at the same frequency in the backward direction (left side of Fig. 1c), we find a modified phase-matching equation:

$$q(\Omega) - \Delta q_{\text{nr}} = k_+(\omega_{\text{f}} - \Omega) - k_-(\omega_{\text{f}}). \tag{11}$$

Here a wavevector shift term $\Delta q_{\text{nr}}$ is introduced. This originates from the traveling acoustic wave which breaks symmetry between forward- and backward-propagating optical waves; however, we will see that its magnitude depends only on the group velocities of the optical waves and the Stokes frequency shift. We can calculate this wavevector by subtracting the two conditions:

$$\Delta q_{\text{nr}} = k_+(\omega_{\text{f}}) - k_+(\omega_{\text{f}} - \Omega) + k_-(\omega_{\text{f}}) - k_-(\omega_{\text{f}} - \Omega) \tag{12}$$

Once again assuming linear dispersion around the operating bandwidth, this term can be approximated as

$$\Delta q_{\text{nr}} = \frac{\partial k_+}{\partial \omega}(\Omega) + \frac{\partial k_-}{\partial \omega}(\Omega) = \frac{n_{\text{g},+} + n_{\text{g},-}}{c}\Omega. \tag{13}$$

When light propagates in the backward direction, scattered Stokes light accumulates a relative phase mismatch $\Delta q_{\text{nr}} L$, where $L$ is the device length. Provided that $\Delta q_{\text{nr}} L \gg 1$, the modulation process will not be phase-matched. This is the necessary condition for nonreciprocal operation.

Interestingly, backward-propagating light may be phase-matched to an inter-band scattering process at a nearby frequency $\omega_{\text{b}}$ when the nonreciprocal wavevector mismatch $\Delta q_{\text{nr}}$ is cancelled by the dispersive wavevector mismatch $\Delta q_{\text{pm}}$. This results in a typical forward/backward modulation response diagrammed in Fig. 1d. We can calculate the frequency difference between $\omega_{\text{b}}$ and $\omega_{\text{f}}$ by requiring that $\Delta q_{\text{nr}} = \Delta q_{\text{pm}}$:

$$\frac{n_{\text{g},+} - n_{\text{g},-}}{c}(\omega_{\text{f}} - \omega_{\text{b}}) = \frac{n_{\text{g},+} + n_{\text{g},-}}{c}\Omega. \tag{14}$$

This simplifies to:

$$\omega_{\text{f}} - \omega_{\text{b}} = \frac{n_{\text{g},+} + n_{\text{g},-}}{n_{\text{g},+} - n_{\text{g},-}}\Omega. \tag{15}$$

This splitting is larger when the dispersion curves for the two modes are more nearly parallel, so that an appreciable frequency difference for light is required to supply the necessary phase mismatch. For our devices with measured group indices $n_{\text{g},+} = 4.0595$ and $n_{\text{g},-} = 4.1853$ [5], and $\Omega_{\text{B}} = 2\pi 5.70$ GHz, $\Delta\omega \approx 65.5\Omega_{\text{B}} = 2\pi 374$ GHz. This corresponds to a 3 nm wavelength-splitting at an optical probe wavelength of 1540 nm, which agrees very well with measured data.

### C. Essential Condition for Significant Nonreciprocity

We have seen that both the phase-matching bandwidth and nonreciprocal frequency splitting for the NIBS process scale inversely with the difference of optical group indices. Therefore, reducing this difference directly increases the bandwidth of operation, and also increases the frequency-splitting between forward and backward phase-matching. For a significant nonreciprocal response to occur, however, it is ideal to have the splitting between forward and backward modulation frequencies be much larger than the operation bandwidth. Here we briefly derive a general characteristic length for this condition to be satisfied.

To have a large frequency-splitting to bandwidth ratio, we require that the half-width at half-maximum (HWHM) bandwidth of the modulator response is much smaller than $|\omega_\mathrm{f} - \omega_\mathrm{b}|$:

$$\frac{2 \cdot 1.39 c}{L} \frac{1}{|n_{\mathrm{g},+} - n_{\mathrm{g},-}|} \ll \frac{n_{\mathrm{g},+} + n_{\mathrm{g},-}}{|n_{\mathrm{g},+} - n_{\mathrm{g},-}|} \Omega \tag{16}$$

which gives a fundamental length scale for "good" nonreciprocity to occur:

$$L \gg \frac{2.78 c}{\Omega \left( n_{\mathrm{g},+} + n_{\mathrm{g},-} \right)} \tag{17}$$

which is 2.8 mm for the silicon waveguides used in the NIBS modulator. The current-generation devices have lengths of 2.4 cm, which seems to satisfy this condition reasonably well, even in the presence of fabrication inhomogeneities. Note that this condition necessitates the use of either a large frequency shift $\Omega$, or a long device length $L$ to achieve nonreciprocity through traveling-wave inter-band photonic transitions of this type.

### D. Effect of Different Waveguide Core Sizes on Phase-Matching

To inhibit optical cross-talk, the drive and modulator waveguides of the NIBS modulator used in the main text are designed to have different core widths. As a result, the phonon mode generated in the drive waveguide at optical wavelength $\lambda_\mathrm{d}$ phase matches to inter-band scattering in the modulator waveguide at a disparate wavelength $\lambda_\mathrm{m}$. The relationship between these two wavelengths can be determined through the phase-matching requirement for the Brillouin process:

$$\frac{2\pi}{\lambda_\mathrm{d}} \left( n_{\mathrm{p},+}^{(1)} (\lambda_\mathrm{d}) - n_{\mathrm{p},-}^{(1)} (\lambda_\mathrm{d}) \right) = \frac{2\pi}{\lambda_\mathrm{m}} \left( n_{\mathrm{p},+}^{(2)} (\lambda_\mathrm{m}) - n_{\mathrm{p},-}^{(2)} (\lambda_\mathrm{m}) \right). \tag{18}$$

Here $n_{\mathrm{p},+}^{(1)}$ and $n_{\mathrm{p},+}^{(2)}$ are the phase indices for the symmetric modes of the drive and modulator waveguides, respectively, and $n_{\mathrm{p},-}^{(1)}$ and $n_{\mathrm{p},-}^{(2)}$ are the phase indices for the anti-symmetric modes. This condition can be written more succinctly as

$$\frac{\lambda_\mathrm{m}}{\lambda_\mathrm{e}} = \frac{n_{\mathrm{p},+}^{(2)} - n_{\mathrm{p},-}^{(2)}}{n_{\mathrm{p},+}^{(1)} - n_{\mathrm{p},-}^{(1)}}. \tag{19}$$

This condition can be used to design devices which operate across very different wavelength bands. For example, by designing a dual-core NIBS modulator with drive core width $w = 1.5$ μm ($n_{p,+}^{(1)} - n_{p,-}^{(1)} = 0.112$ at $\lambda = 1550$ nm) and modulator core width $w = 2.18$ μm ($n_{p,+}^{(2)} - n_{p,-}^{(2)} = 0.112$ at $\lambda = 2100$ nm), modulation can in principle be driven using optical waves >500 nm away from the probe wavelength.

### E. Phase-mismatched Lineshape

When the NIBS scattering process is perfectly phase-matched, the frequency response of modulation efficiency gives the expected Lorentzian-like lineshape determined by the lifetime of the resonant phonon mode. However, if the probe wave is slightly detuned from the ideal wavelength for phase-matching, then the scattered light accumulates a frequency-dependent phase mismatch relative to the probe according to Eq. 7 as it traverses the device. In this case, the frequency response of the modulation efficiency can take on many new shapes, including asymmetric lineshapes, sharp frequency rolloffs, and notch-like features. Several of these lineshapes are plotted as a function of wavevector mismatch (probe wavelength) in Fig. 2.

Although inhomogeneities in device fabrication complicate the exact behavior of phase-matching in these devices, all of these lineshapes can be reproduced using a simple model that includes (1) a constant wavevector mismatch $\Delta q_{pm}$ along the device and (2) a change in Brillouin frequency along the device length. The latter is known to occur in nanoscale Brillouin devices, resulting in broadening of the resonance lineshape [6], but plays an additional role here.

Let the amplitudes of the drive-waveguide optical waves be $a_p^{(1)}$ and $a_s^{(1)}$ and the amplitude of the probe wave in the modulator waveguide be $a_p^{(2)}$. Then the spatial evolution of the amplitude of the scattered Stokes wave $a_s^{(2)}$ can be described by the differential equation

$$\frac{\partial a_s^{(2)}}{\partial z} = e^{i\Delta q_{pm} z} a_p^{(1)}(z) a_s^{(1)*}(z) a_p^{(2)}(z) \frac{\gamma_B(z) \Gamma/2}{\Omega_B(z) - \Omega - i\Gamma/2}. \tag{20}$$

Here $\gamma_B(z)$ is the nonlinear coupling coefficient and $\Omega_B(z)$ is the phonon resonance frequency, both of which which may vary along the device length, and $\Gamma$ is the intrinsic phonon lifetime.

With various choices for $\gamma_B(z)$ and $\Omega_B(z)$ this equation reproduces most of the interesting frequency response characteristics that are experimentally observed. It should be noted that in reality $\Delta q_{pm}$ is likely also position-dependent since the optical group indices will change in response to small variations in waveguide core size.

Note that in the absence of $z$-dependent inhomogeneities and assuming undepeleted pump fields, Eq. 20 gives the expected sinc-like modulation response as wavevector detuning is changed:

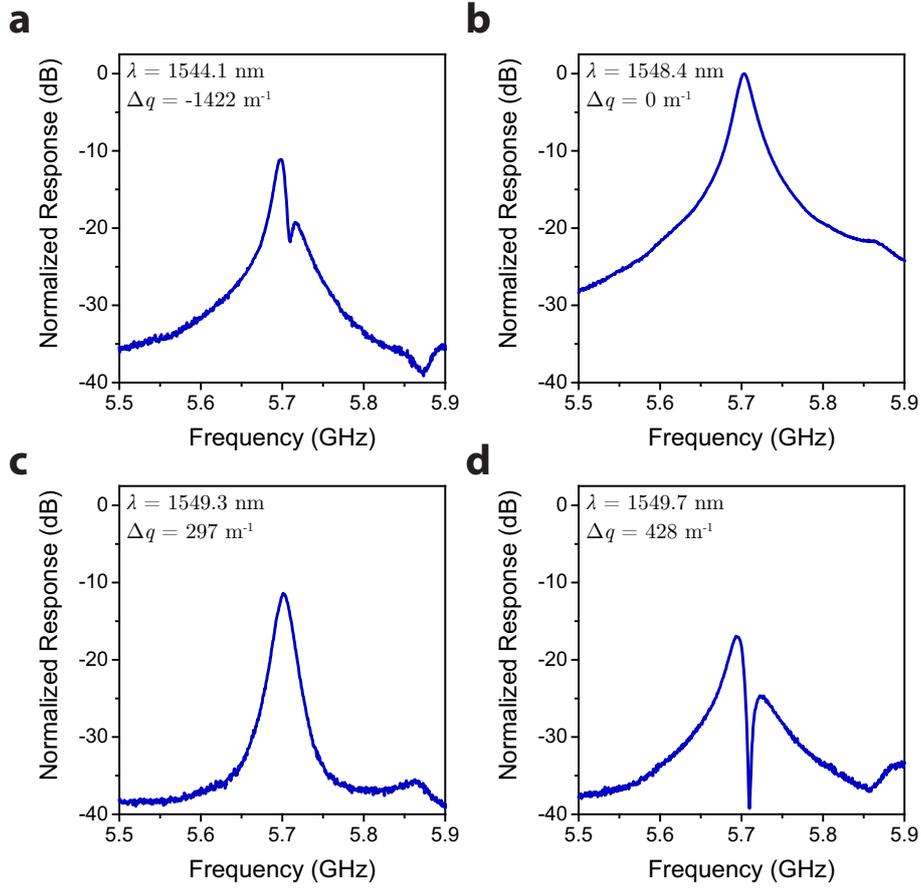

FIG. 2: Frequency response of the NIBS modulator as the probe wavelength, and hence wavevector mismatch, is varied (a) $\lambda_{\rm p} = 1544.1$ nm, $\Delta q = -1422$ m$^{-1}$ (b) $\lambda_{\rm p} = 1548.4$ nm, $\Delta q = 0$ m$^{-1}$ ("phase-matched") (c) $\lambda_{\rm p} = 1549.3$ nm, $\Delta q = 297$ m$^{-1}$ (d) $\lambda_{\rm p} = 1549.7$ nm, $\Delta q = 428$ m$^{-1}$.

$$a_{\rm s}^{(2)}(L) = a_{\rm p}^{(1)} a_{\rm s}^{(1)*} a_{\rm p}^{(2)} \frac{\gamma_{\rm B}\Gamma/2}{\Omega_{\rm B} - \Omega - i\Gamma/2} \int_0^L e^{i\Delta q_{\rm pm} z} dz \\ = a_{\rm p}^{(1)} a_{\rm s}^{(1)*} a_{\rm p}^{(2)} \frac{\gamma_{\rm B}\Gamma/2}{\Omega_{\rm B} - \Omega - i\Gamma/2} L e^{i\Delta q_{\rm pm} L/2} {\rm sinc}\left(\frac{\Delta q_{\rm pm} L}{2}\right). \quad (21)$$

Then the output modulation signal power is given by

$$P_{\rm s}^{(2)}(L) = a_{\rm s}^{(2)}(L) a_{\rm s}^{(2)*}(L) = P_{\rm p}^{(1)} P_{\rm s}^{(1)} P_{\rm p}^{(2)} L^2 \frac{|\gamma_{\rm B}|^2 \Gamma^2/4}{(\Omega_{\rm B} - \Omega)^2 + \Gamma^2/4} {\rm sinc}^2\left(\frac{\Delta q_{\rm pm} L}{2}\right). \quad (22)$$

### F. Improving Modulator Bandwidth with Dispersion Engineering

The phase matching bandwidth of the NIBS process is determined by the group velocities of the optical modes, as described in Section IA, and is given by Eq. (7). Specifically, this bandwidth is inversely proportional to the difference in group indexes of the optical modes $\Delta n_{\rm g} = |n_{\rm g,+} - n_{\rm g,-}|$. Reducing the difference in group indexes will therefore enhance the bandwidth of the nonreciprocal modulator for a device of a given length $L$.

99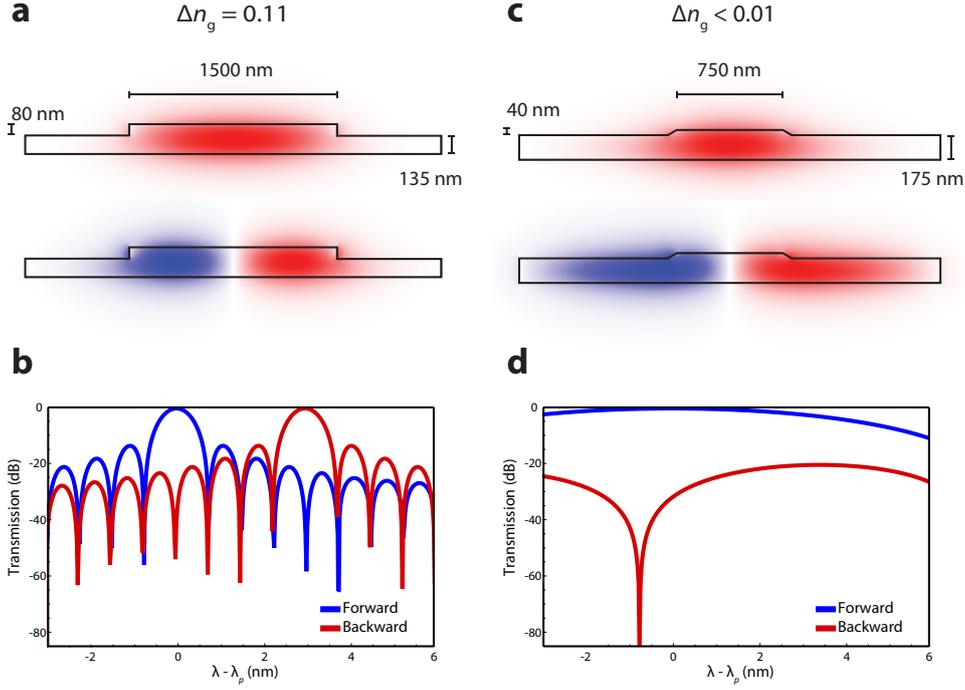

FIG. 3: (a) Dimensions of the ridge waveguides used in NIBS measurements, and simulated $x$ component of the electric field for the first two optical modes supported by the waveguide. (b) Calculated normalized transmission as a function of wavelength shows the phase-matching sinc-squared response in the forward and backwards directions. (c) An alternative ridge waveguide design with reduced ridge dimensions and angled sidewalls, resulting in a tenfold reduction in the difference of group indexes. (d) Calculated normalized transmission of a NIBS process using waveguides shown in (c). The reduction of $\Delta n_g$ increases the bandwidth by a factor of 10.

Furthermore, this enhancement in bandwidth does not affect the nonreciprocal performance, as the frequency splitting between forward and backward phase-matching also scales inversely with $\Delta n_g$ (Section IC).

The optical dispersion properties of the waveguides are determined by the refractive index profile of the waveguide cross section, and by the waveguide geometry. Therefore, minimizing $\Delta n_g$ can be achieved by modifying the waveguide design and material properties [7–12] to maximize the phase-matched bandwidth.

As an example, we compare the bandwidth of a device similar to the one measured in this study and an alternative device with a modified waveguide geometry. Fig. 3a shows the optical waveguide design used for both waveguide cores of the NIBS modulator from the main text, and the two first optical mode profiles as simulated by a finite-element mode solver. This ridge waveguide device has a simulated difference of group indexes $\Delta n_g = 0.11$, in good agreement with measurements. The expected transmission response of this NIBS device in the forward and backward directions is illustrated in Fig. 3b. An alternative, weakly-guiding ridge design is illustrated in Fig. 3c, where the dimensions of the guiding ridge are half of those in (a) and have a 65° angle for the ridge sidewalls. The difference in group indexes of this design is calculated to be $\Delta n_g = 0.01$, a tenfold reduction compared to the current device design. The transmission response of the modified device is

illustrated in Fig. 3d, showing a corresponding tenfold increase in bandwidth. Similar results can be achieved by a variety of different design modifications to enable ultra-broad bandwidth nonreciprocal devices in silicon photonic circuits.

## II. COUPLED AMPLITUDE EQUATIONS AND ENERGY TRANSFER DYNAMICS

In this section, we write down the coupled amplitude equations which describe the spatial evolution of optical and acoustic fields within the optically-driven NIBS modulator. Throughout this discussion, we assume a two-waveguide system, with each waveguide core guiding pump and Stokes waves in two separate optical modes. However, this treatment may be extended to more general systems such as polarization- or spatially-multiplexed optical fiber, or light fields of disparate wavelengths or spatial modes within the same Brillouin-active waveguide core, provided that inter-mode linear and nonlinear crosstalk is negligible.

### A. General Coupling Equations Including Nonlinear Loss

We begin with the case of on-resonant interaction in the steady state, where we have assumed phase matching and that all optical frequencies are approximately equal for purposes of energy conservation. In the drive waveguide, we inject two guided optical waves at frequencies $\omega_p^{(1)}$ and $\omega_s^{(1)} = \omega_p^{(1)} - \Omega$ with amplitudes $a_p^{(1)}$ and $a_s^{(1)}$. When these waves are coupled through a common phonon field with amplitude $b$, their coupled-amplitude equations of motion are [13, 14]

$$\frac{\partial a_p^{(1)}}{\partial z} = -\frac{G^{(1)}}{2}a_s^{(1)}b \; - \; \frac{1}{2}\left(\alpha_p^{(1)} + \beta_{pp}^{(1)}\left|a_p^{(1)}\right|^2 + \gamma_{ppp}^{(1)}\left|a_p^{(1)}\right|^4\right)a_p^{(1)} -$$
$$\frac{1}{2}\left(2\beta_{ps}^{(1)} + 4\gamma_{pps}^{(1)}\left|a_p^{(1)}\right|^2 + \gamma_{pss}^{(1)}\left|a_s^{(1)}\right|^2\right)\left|a_s^{(1)}\right|^2 a_p^{(1)} \quad (23)$$

$$\frac{\partial a_s^{(1)}}{\partial z} = \frac{G^{(1)}}{2}a_p^{(1)}b^* \; - \; \frac{1}{2}\left(\alpha_s^{(1)} + \beta_{ss}^{(1)}\left|a_s^{(1)}\right|^2 + \gamma_{sss}^{(1)}\left|a_s^{(1)}\right|^4\right)a_s^{(1)} -$$
$$\frac{1}{2}\left(2\beta_{sp}^{(1)} + 4\gamma_{ssp}^{(1)}\left|a_s^{(1)}\right|^2 + \gamma_{spp}^{(1)}\left|a_p^{(1)}\right|^2\right)\left|a_p^{(1)}\right|^2 a_s^{(1)} \quad (24)$$

where we have assumed that the phonon field is spatially heavily damped compared to the distance over which appreciable optical energy transfer occurs. In this case the phonon field follows the spatial evolution of the optical fields and can be written as:

$$b = a_s^{(1)*}a_p^{(1)}. \quad (25)$$

In these equations, $G^{(1)}$ is the real-valued Brillouin coupling coefficient, $\alpha_i$ is the linear power loss coefficient for mode $i$, $\beta_{ii}$ and $\beta_{ij}$ are the intra- and inter-modal nonlinear loss coefficients due to two-photon absorption

(TPA). $\gamma_{iii}$ is the intra-modal nonlinear loss coefficient for TPA-induced free carrier absorption (FCA), while $\gamma_{ijj}$ and $\gamma_{iij}$ are the inter-modal FCA loss coefficients. Here $i$ and $j$ are dummy indices which refer to either optical field (mode).

The optical amplitudes are normalized such that $P_p^{(1)}(z) = \left|a_p^{(1)}\right|^2$ and $P_s^{(1)}(z) = \left|a_s^{(1)}\right|^2$ and the phonon field is normalized such that $P_b = \frac{\Omega_B}{\omega_s^{(1)}} \frac{v_{b,g}}{\Gamma_B} G^{(1)} |b|^2$, where $v_{b,g}$ is the acoustic group velocity and $\Gamma_B$ is the acoustic decay rate.

Equations (1)-(3) describe inter-modal Brillouin coupling in the presence of nonlinear loss for two optical modes guided in the same waveguide coupled though a single phonon field.

Next, we modify and extend these equations of motion to describe the NIBS process by including two additional optical waves guided in a separate waveguide which couple to the same acoustic phonon mode. In general these fields, with amplitudes $a_p^{(2)}$ and $a_s^{(2)}$, will be at a distinct set of optical frequencies separated by the phonon frequency $\omega_s^{(2)} = \omega_p^{(2)} - \Omega$. Note that depending on the initial conditions and phase-matching configuration of the system, we can have energy transfer in either direction between these two fields. In general, the motion of these fields is governed by equations structurally identical to Eqs. (1)-(2):

$$\frac{\partial a_p^{(2)}}{\partial z} = -\frac{G^{(2)}}{2} a_s^{(2)} b \; - \; \frac{1}{2}\left(\alpha_p^{(2)} + \beta_{pp}^{(2)}\left|a_p^{(2)}\right|^2 + \gamma_{ppp}^{(2)}\left|a_p^{(2)}\right|^4\right)a_p^{(2)} - \\ \frac{1}{2}\left(2\beta_{ps}^{(2)} + 4\gamma_{pps}^{(2)}\left|a_p^{(2)}\right|^2 + \gamma_{pss}^{(2)}\left|a_s^{(2)}\right|^2\right)\left|a_s^{(2)}\right|^2 a_p^{(2)} \quad (26)$$

$$\frac{\partial a_s^{(2)}}{\partial z} = \frac{G^{(2)}}{2} a_p^{(2)} b^* \; - \; \frac{1}{2}\left(\alpha_s^{(2)} + \beta_{ss}^{(2)}\left|a_s^{(2)}\right|^2 + \gamma_{sss}^{(2)}\left|a_s^{(2)}\right|^4\right)a_s^{(2)} - \\ \frac{1}{2}\left(2\beta_{sp}^{(2)} + 4\gamma_{ssp}^{(2)}\left|a_s^{(2)}\right|^2 + \gamma_{spp}^{(2)}\left|a_p^{(2)}\right|^2\right)\left|a_p^{(2)}\right|^2 a_s^{(2)} \quad (27)$$

Where we must also modify the phonon field to include driving terms from both waveguides:

$$b = a_s^{(1)*} a_p^{(1)} + a_s^{(2)*} a_p^{(2)}. \quad (28)$$

Note that this opens the door to the possibility of action on the phonon field by the modulator waveguide. In the case that the amplitude product between the two terms is different in sign, this can lead to destructive interference between the two driving terms (i.e. a steady-state) in the regime of strongly-coupled dynamics.

Depending on geometric asymmetries between drive and modulator waveguides, the linear and nonlinear coefficients can be different between the two waveguides. In this present work, drive and modulator waveguides are almost identical and symmetric, so $G^{(1)} \approx G^{(2)}$, $\alpha_i^{(1)} \approx \alpha_i^{(2)}$, $\beta_{ij}^{(1)} \approx \beta_{ij}^{(2)}$, and $\gamma_{ijk}^{(1)} \approx \gamma_{ijk}^{(2)}$. We have also neglected the potential for inter-core nonlinear loss, for example that arising from diffusion of free carriers from one core to the other. We do not observe any excess inter-core loss even at the highest tested powers throughout our experiments, so this seems to be a good approximation.

Together, Eqs. (23-24) and (26-28) describe the general dynamics of the NIBS modulator studied in this work and are used to calculate the theoretical trend in Fig. 7.

## B. Approximate Analytic Solution to Coupling Equations

In order to understand the behavior of and ultimate limits to energy transfer through the NIBS process, we next seek a closed-form solution to coupling equations of the type of Section IA. We neglect nonlinear loss, which we can later re-introduce as a modification to a single linear loss parameter $\alpha_s^{(1)} = \alpha_p^{(1)} \equiv \alpha^{(1)}$. We also neglect the action of the modulator waveguide fields on the phonon amplitude $\left| a_s^{(1)} a_p^{(1)} \right| \gg \left| a_s^{(2)} a_p^{(2)} \right|$. In this case, the equations of motion for the five field amplitudes are:

$$\frac{\partial a_p^{(1)}}{\partial z} = -\frac{G^{(1)}}{2} a_s^{(1)} b - \frac{1}{2} \alpha^{(1)} a_p^{(1)} \tag{29}$$

$$\frac{\partial a_s^{(1)}}{\partial z} = \frac{G^{(1)}}{2} a_p^{(1)} b^* - \frac{1}{2} \alpha^{(1)} a_s^{(1)} \tag{30}$$

$$\frac{\partial a_p^{(2)}}{\partial z} = -\frac{G^{(2)}}{2} a_s^{(2)} b - \frac{1}{2} \alpha^{(2)} a_p^{(2)} \tag{31}$$

$$\frac{\partial a_s^{(2)}}{\partial z} = \frac{G^{(2)}}{2} a_p^{(2)} b^* - \frac{1}{2} \alpha^{(2)} a_s^{(2)} \tag{32}$$

$$b = a_s^{(1)*} a_p^{(1)}. \tag{33}$$

Substituting Eq. (33) into Eqs. (29)-(30) allows us to write the coupled equations for the two fields in the drive waveguide as

$$\frac{\partial a_p^{(1)}}{\partial z} = -\frac{G^{(1)}}{2} \left| a_s^{(1)} \right|^2 a_p^{(1)} - \frac{\alpha^{(1)}}{2} a_p^{(1)} \tag{34}$$

$$\frac{\partial a_s^{(1)}}{\partial z} = \frac{G^{(1)}}{2} \left| a_p^{(1)} \right|^2 a_s^{(1)} - \frac{\alpha^{(1)}}{2} a_s^{(1)} \tag{35}$$

Since these equations are decoupled from those of the modulator waveguide, we first seek the general solution to their dynamics, which will allow us to write down the spatial evolution of the phonon field $b$.

To simplify these equations, we make a change of variables $a_p^{(1)} = e^{-\alpha^{(1)}z/2} q_p^{(1)}$, $a_s^{(1)} = e^{-\alpha^{(1)}z/2} q_s^{(1)}$. The transformed equations read

$$\frac{\partial q_p^{(1)}}{\partial z} = -\frac{G^{(1)}}{2} \left| q_s^{(1)} \right|^2 e^{-\alpha^{(1)}z} q_p^{(1)} \tag{36}$$

$$\frac{\partial q_s^{(1)}}{\partial z} = \frac{G^{(1)}}{2} \left| q_p^{(1)} \right|^2 e^{-\alpha^{(1)}z} q_s^{(1)} \tag{37}$$

Note that these equations satisfy the conservation relation $\frac{\partial}{\partial z}\left(q_{\text{p}}^{(1)*}q_{\text{p}}^{(1)} + q_{\text{s}}^{(1)*}q_{\text{s}}^{(1)}\right) = 0$. As a result, $\left|q_{\text{p}}^{(1)}\right|^2 + \left|q_{\text{s}}^{(1)}\right|^2$ is a constant equal to the total input power $P_{\text{in}}^{(1)}$. This allows us to rewrite Eqs. 14-15 as

$$\frac{\partial q_{\text{p}}^{(1)}}{\partial z} = -\frac{G^{(1)}}{2}\left(P_{\text{in}}^{(1)} - \left|q_{\text{p}}^{(1)}\right|^2\right)e^{-\alpha^{(1)}z}q_{\text{p}}^{(1)}, \tag{38}$$

$$\frac{\partial q_{\text{s}}^{(1)}}{\partial z} = \frac{G^{(1)}}{2}\left(P_{\text{in}}^{(1)} - \left|q_{\text{s}}^{(1)}\right|^2\right)e^{-\alpha^{(1)}z}q_{\text{s}}^{(1)}. \tag{39}$$

We need to make one more observation to solve these (now decoupled) equations. Note that, while in general $q_{\text{p}}^{(1)}$ and $q_{\text{s}}^{(1)}$ are complex numbers, their complex phase is unchanged with propagation. In other words, we can make another set of substitutions $r_{\text{p}}^{(1)} = e^{-i\phi_{\text{p}}^{(1)}}q_{\text{p}}^{(1)}$ and $r_{\text{s}}^{(1)} = e^{-i\phi_{\text{s}}^{(1)}}q_{\text{s}}^{(1)}$, where $r_{\text{p}}^{(1)}$ and $r_{\text{s}}^{(1)}$ are real and $e^{i\phi_{\text{p}}^{(1)}}$ and $e^{i\phi_{\text{s}}^{(1)}}$ are the phase factors of the pump and Stokes waves, respectively. The equations governing the propagation of these real variables are

$$\frac{\partial r_{\text{p}}^{(1)}}{\partial z} = -\frac{G^{(1)}}{2}\left(P_{\text{in}}^{(1)} - \left(r_{\text{p}}^{(1)}\right)^2\right)e^{-\alpha^{(1)}z}r_{\text{p}}^{(1)}, \tag{40}$$

$$\frac{\partial r_{\text{s}}^{(1)}}{\partial z} = \frac{G^{(1)}}{2}\left(P_{\text{in}}^{(1)} - \left(r_{\text{s}}^{(1)}\right)^2\right)e^{-\alpha^{(1)}z}r_{\text{s}}^{(1)} \tag{41}$$

and each is separable with the solutions:

$$r_{\text{p}}^{(1)2}(z) = \frac{P_{\text{in}}^{(1)}e^{\frac{G^{(1)}P_{\text{in}}^{(1)}}{\alpha^{(1)}}\left(e^{-\alpha^{(1)}z}-1\right)}}{k + e^{\frac{G^{(1)}P_{\text{in}}^{(1)}}{\alpha^{(1)}}\left(e^{-\alpha^{(1)}z}-1\right)}} \tag{42}$$

$$r_{\text{s}}^{(1)2}(z) = \frac{P_{\text{in}}^{(1)}k}{k + e^{\frac{G^{(1)}P_{\text{in}}^{(1)}}{\alpha^{(1)}}\left(e^{-\alpha^{(1)}z}-1\right)}} \tag{43}$$

where $k \equiv P_{\text{s}}^{(1)}(z=0)/P_{\text{p}}^{(1)}(z=0)$ is the ratio of input Stokes to pump powers. Since the exponential terms, $k$, and $P_{\text{in}}^{(1)}$ are all positive, we can take the positive roots and transform back to field amplitudes using $a_{\text{p}}^{(1)} = e^{i\phi_{\text{p}}^{(1)}}e^{-\alpha^{(1)}z/2}r_{\text{p}}^{(1)}$, $a_{\text{s}}^{(1)} = e^{i\phi_{\text{s}}^{(1)}}e^{-\alpha^{(1)}z/2}r_{\text{s}}^{(1)}$:

$$a_{\text{p}}^{(1)}(z) = \frac{e^{i\phi_{\text{p}}^{(1)}}\sqrt{P_{\text{in}}^{(1)}}e^{-\alpha^{(1)}z/2}e^{\frac{G^{(1)}P_{\text{in}}^{(1)}}{2\alpha^{(1)}}\left(e^{-\alpha^{(1)}z}-1\right)}}{\sqrt{k + e^{\frac{G^{(1)}P_{\text{in}}^{(1)}}{\alpha^{(1)}}\left(e^{-\alpha^{(1)}z}-1\right)}}} \tag{44}$$

$$a_{\text{s}}^{(1)}(z) = \frac{e^{i\phi_{\text{s}}^{(1)}}e^{-\alpha^{(1)}z/2}\sqrt{P_{\text{in}}^{(1)}k}}{\sqrt{k + e^{\frac{G^{(1)}P_{\text{in}}^{(1)}}{\alpha^{(1)}}\left(e^{-\alpha^{(1)}z}-1\right)}}} \tag{45}$$

The resulting driven phonon amplitude is

$$b(z) = a_s^{(1)*}(z) a_p^{(1)}(z) = \frac{e^{i\left(\phi_p^{(1)} - \phi_s^{(1)}\right)} e^{-\alpha^{(1)} z} P_{\text{in}}^{(1)} \sqrt{k} e^{\frac{G^{(1)} P_{\text{in}}^{(1)}}{2\alpha^{(1)}}\left(e^{-\alpha^{(1)} z} - 1\right)}}{k + e^{\frac{G^{(1)} P_{\text{in}}^{(1)}}{\alpha^{(1)}}\left(e^{-\alpha^{(1)} z} - 1\right)}} \equiv e^{i\phi_b} |b(z)|. \quad (46)$$

Here we have also rewritten the complex phonon amplitude as consisting of a complex phase $e^{i\phi_b} = e^{i\left(\phi_p^{(1)} - \phi_s^{(1)}\right)}$ and a real amplitude, $|b(z)|$.

We now return to the Eqs. (9)-(10) for the modulator-waveguide optical field amplitudes and make the change of variables $a_p^{(2)} = e^{-\alpha^{(2)} z/2} q_p^{(2)}$, $a_s^{(2)} = e^{-\alpha^{(2)} z/2} q_s^{(2)}$ to eliminate the optical loss term:

$$\frac{\partial q_p^{(2)}}{\partial z} = -\frac{G^{(2)}}{2} q_s^{(2)} b = -\frac{G^{(2)}}{2} q_s^{(2)} e^{i\phi_b} |b(z)| \quad (47)$$

$$\frac{\partial q_s^{(2)}}{\partial z} = \frac{G^{(2)}}{2} q_p^{(2)} b^* = \frac{G^{(2)}}{2} q_p^{(2)} e^{-i\phi_b} |b(z)| \quad (48)$$

As before, we seek to transform these differential equations in complex variables to a set of purely real variables. We use the substitutions $r_p^{(2)} = e^{-i\phi_p^{(2)}} q_p^{(2)}$ and $r_s^{(2)} = e^{-i\left(\phi_b + \phi_p^{(2)}\right)} q_s^{(2)}$, where $e^{i\phi_p^{(2)}}$ is the input phase of the pump wave in the modulator waveguide. This transformation eliminates the complex phase of the phonon field to yield the coupled equations

$$\frac{\partial r_p^{(2)}}{\partial z} = -\frac{G^{(2)}}{2} r_s^{(2)} |b(z)| \quad (49)$$

$$\frac{\partial r_s^{(2)}}{\partial z} = \frac{G^{(2)}}{2} r_p^{(2)} |b(z)| \quad (50)$$

Note that while we have assumed an arbitrary phase factor $e^{i\phi_p^{(2)}}$ for the pump wave, we do not have this degree of freedom for the scattered Stokes wave. If we have incident Stokes light that is out-of-phase with the scattered Stokes light (at phase $\phi_b + \phi_p^{(2)}$), then the dynamics of this problem become more complex and cannot easily be uncoupled. Here we assume that we have no incident Stokes light, as in the case of typical operation. Then we may take $r_p^{(2)}$ to be real since we have already factored out an arbitrary phase, and hence $r_s^{(2)}$ will also be real.

These equations then satisfy the conservation relation $\frac{\partial}{\partial z}\left(\left(r_p^{(2)}\right)^2 + \left(r_s^{(2)}\right)^2\right) = 0$, so we can write $\left(r_p^{(2)}\right)^2 + \left(r_s^{(2)}\right)^2 = P_{\text{in}}^{(2)}$, where $P_{\text{in}}^{(2)}$ is the total incident power in the drive waveguide and is assumed to be incident entirely in the pump wave, i.e. $P_{\text{in}}^{(2)} = \left(r_p^{(2)}\right)^2 (z=0)$. The equation governing the spatial evolution of the Stokes wave becomes

$$\frac{\partial r_s^{(2)}}{\partial z} = \frac{G^{(2)}}{2} \sqrt{P_{\text{in}}^{(2)} - \left(r_s^{(2)}\right)^2} |b(z)| \quad (51)$$

This equation is again separable as:

$$\frac{\partial r_\text{s}^{(2)}}{\sqrt{P_\text{in}^{(2)} - \left(r_\text{s}^{(2)}\right)^2}} = \frac{G^{(2)}}{2} \left|b(z)\right| \partial z \tag{52}$$

In other words, for any NIBS modulation process, provided that we can integrate the driven phonon field over space, we can find an expression for the Stokes signal power. Here this equation becomes

$$\int \frac{\partial r_\text{s}^{(2)}}{\sqrt{P_\text{in}^{(2)} - \left(r_\text{s}^{(2)}\right)^2}} = \int \frac{G^{(2)}}{2} \frac{e^{-\alpha^{(1)}z} P_\text{in}^{(1)} \sqrt{k} e^{\frac{G^{(1)} P_\text{in}^{(1)}}{2\alpha^{(1)}}\left(e^{-\alpha^{(1)}z}-1\right)}}{k + e^{\frac{G^{(1)} P_\text{in}^{(1)}}{\alpha^{(1)}}\left(e^{-\alpha^{(1)}z}-1\right)}} \partial z \tag{53}$$

The righthand side is integrable with the substitution $u = e^{-\alpha^{(1)}z}$.

$$\tan^{-1}\left(\frac{r_\text{s}^{(2)}}{\sqrt{P_\text{in}^{(2)} - \left(r_\text{s}^{(2)}\right)^2}}\right) = \frac{G^{(2)}}{G^{(1)}} \left(\tan^{-1}\left(\frac{1}{\sqrt{k}}\right) - \tan^{-1}\left(\frac{e^{\frac{G^{(1)}P}{2\alpha_1}\left(e^{-\alpha^{(1)}z}-1\right)}}{\sqrt{k}}\right)\right) \tag{54}$$

which simplifies to

$$r_\text{s}^{(2)} = \sqrt{P_\text{in}^{(2)}} \sin\left(\frac{G^{(2)}}{G^{(1)}}\left(\tan^{-1}\left(\frac{1}{\sqrt{k}}\right) - \tan^{-1}\left(\frac{e^{\frac{G^{(1)}P_\text{in}^{(1)}}{2\alpha_1}\left(e^{-\alpha^{(1)}z}-1\right)}}{\sqrt{k}}\right)\right)\right). \tag{55}$$

Substituting back, we find the complex amplitude using $a_\text{s}^{(2)} = e^{i\left(\phi_\text{p}^{(1)} - \phi_\text{s}^{(1)} + \phi_\text{p}^{(2)}\right)} e^{-\alpha^{(2)}z/2} r_\text{s}^{(2)}$:

$$a_\text{s}^{(2)} = \sqrt{P_\text{in}^{(2)}} e^{i\left(\phi_\text{p}^{(1)} - \phi_\text{s}^{(1)} + \phi_\text{p}^{(2)}\right)} e^{-\alpha^{(2)}z/2} \sin\left(\frac{G^{(2)}}{G^{(1)}}\left(\tan^{-1}\left(\frac{1}{\sqrt{k}}\right) - \tan^{-1}\left(\frac{e^{\frac{G^{(1)}P_\text{in}^{(1)}}{2\alpha_1}\left(e^{-\alpha^{(1)}z}-1\right)}}{\sqrt{k}}\right)\right)\right). \tag{56}$$

This equation describes the spatial evolution of the scattered Stokes amplitude in the modulator waveguide as a function of the other three incident fields, the Brillouin couplings in each waveguide, and propagation losses. We define the total modulation efficiency $\eta^2$ as the output scattered light power relative to the incident power in the modulator waveguide:

$$\eta^2 \equiv \frac{P_\text{s}^{(2)}(L)}{P_\text{in}^{(2)}} = \frac{a_\text{s}^{(2)*}(L) a_\text{s}^{(2)}(L)}{P_\text{in}^{(2)}}$$

$$= e^{-\alpha^{(2)}z} \sin^2\left(\frac{G^{(2)}}{G^{(1)}}\left(\tan^{-1}\left(\frac{1}{\sqrt{k}}\right) - \tan^{-1}\left(\frac{e^{\frac{G^{(1)}P_\text{in}^{(1)}}{2\alpha_1}\left(e^{-\alpha^{(1)}L}-1\right)}}{\sqrt{k}}\right)\right)\right). \tag{57}$$

Where we have taken $z = L$ to be the total device length. For maximum efficiency to occur, the expression inside the sine-squared term should be equal to $\pi/2$.

For given values of $G^{(1)}$ and $P_{\text{in}}^{(1)}$, (i.e. given a device design and power budget), this expression is maximized when

$$k = e^{\frac{G^{(1)} P_{\text{in}}^{(1)}}{2\alpha^{(1)}} \left(e^{-\alpha^{(1)} L} - 1\right)} \tag{58}$$

In other words, there is an optimal way to bias the relative powers of the two waves in the drive waveguide. Given this optimal power biasing, the minimum pump power to reach unity efficiency (complete power conversion, neglecting linear insertion loss, in the modulator waveguide) is:

$$P_{\text{in}}^{(1)} = \frac{2\alpha^{(1)}}{G^{(1)} \left(e^{-\alpha^{(1)} L} - 1\right)} \log\left(\tan^2\left(\frac{\pi}{4}\left(1 - \frac{G^{(1)}}{G^{(2)}}\right)\right)\right) \tag{59}$$

1. *Special Case: $G^{(1)} = G^{(2)}$*

In symmetrical systems, the Brillouin coupling coefficients for each process are nearly identical. This is the case for the NIBS modulator device studied here where the drive and modulator waveguide core sizes and wavelengths are different by less than 2%. In this situation, the equations governing conversion efficiency simplify dramatically.

In the case where $G^{(1)} = G^{(2)} \equiv G$, Eq. (57) becomes

$$\eta^2 = e^{-\alpha^{(2)} L} \frac{k}{k+1} \frac{\left(e^{\frac{G P_{\text{in}}^{(1)}}{2\alpha^{(1)}}\left(1 - e^{-\alpha^{(1)} L}\right)} - 1\right)^2}{\left(k e^{\frac{G P_{\text{in}}^{(1)}}{\alpha^{(1)}}\left(1 - e^{-\alpha^{(1)} L}\right)} + 1\right)} \tag{60}$$

An absolute upper bound on energy transfer is defined by the relative ratio of input pump to Stokes powers in the modulator waveguide.

$$\lim_{G P_{\text{in}}^{(1)} \to \infty} \eta^2 = e^{-\alpha^{(2)} L} \frac{1}{k+1} \tag{61}$$

In other words, the fraction of power transfer in the modulator waveguide is bounded by the fraction of power transfer in the drive waveguide. This limit results from pump depletion, and hence phonon field attenuation, in this waveguide.

In most realistic systems, there is a practical upper limit on optical power, Brillouin coupling, and device length. In a system where these are fixed, the maximum energy transfer is achieved when the input power ratio $k$ satisfies Eq. (58). When this is the case, the maximum efficiency is given by:

$$\eta_{\text{max}}^2 = e^{-\alpha^{(2)}L} \tanh^2\left(\frac{GP_{\text{in}}^{(1)}\left(1 - e^{-\alpha^{(1)}L}\right)}{4\alpha^{(1)}}\right) \tag{62}$$

If insertion losses are small ($\alpha_1 L \ll 1$), then this expression simplifies further

$$\eta_{\text{max}}^2 = e^{-\alpha^{(2)}L} \tanh^2\left(\frac{GP_{\text{in}}^{(1)}L}{4}\right) \tag{63}$$

### C. Externally Driven Phonon Field

We have derived expressions for energy transfer efficiency given optical pumping of the acoustic phonon mode with injection only at the device input. In order to achieve maximum energy transfer in a small footprint, other acoustic driving schemes may be preferable, e.g. re-injection of pump light along device length, or electromechanical driving of the phonon mode [15, 16]. We next briefly consider the case of an arbitrary phonon amplitude profile.

From Eq. (52), we can derive an analogous result to Eq. (57) for an arbitrary phonon field:

$$\eta^2 \equiv \frac{P_s^{(2)}(L)}{P_{\text{in}}^{(2)}} = e^{-\alpha^{(2)}L} \sin^2\left(\int_0^L \frac{G^{(2)}\left|b(z)\right| dz}{2}\right) \tag{64}$$

If we consider a phonon field $b(z) = b_0$ whose amplitude is constant in space, this expression becomes:

$$\eta^2 = e^{-\alpha^{(2)}L} \sin^2\left(\frac{G^{(2)}b_0 L}{2}\right) \tag{65}$$

This efficiency is maximized when $G^{(2)}b_0 L = \pi$. Since for an optically-driven acoustic wave $b_0 \propto \sqrt{P_p^{(1)}P_s^{(1)}}$, this sets a minimum bound on the optical power necessary to achieve unity modulation efficiency in terms of the total incident power $P_{\text{in}}^{(1)}$; assuming $P_p^{(1)} = P_s^{(1)} = P_{\text{in}}^{(1)}/2$, which locally maximizes the driven phonon amplitude, then $G^{(2)}P_{\text{in}}^{(1)}L > 2\pi$. To practically achieve comparable performance with such gain-power-length products, schemes for re-injection of depleted pump light are necessary. Without such techniques, Eq. (63) gives a condition $G^{(2)}P_{\text{in}}^{(1)}L = 12$ for 99% modulation efficiency in a linear device.

We can calculate the corresponding acoustic power necessary for unity efficiency by invoking the normalization condition $P_b = \frac{\Omega_B}{\omega_s}\frac{v_{b,g}}{\Gamma_B} G \left|b_0\right|^2$ where we assume a single Brillouin coupling coefficient $G$ and single optical Stokes frequency $\omega_s$. Then the power required for complete energy transfer from pump to Stokes waves is:

$$P_b = \frac{\Omega_B}{\omega_s}\frac{v_{b,g}}{\Gamma_B}\frac{\pi^2}{GL^2} \tag{66}$$

which can also be expressed in terms of the distributed optomechanical coupling strength $g_0$ as [13]:

$$P_\text{b} = \hbar\Omega_\text{B} \frac{v_{\text{b,g}} v_\text{s} v_\text{p}}{4|g_0|^2} \frac{\pi^2}{L^2} \tag{67}$$

where $v_\text{s}$ and $v_\text{p}$ are the optical group velocities of the pump and Stokes waves. For a Brillouin-active silicon waveguide with identical parameters to those studied here, this threshold acoustic power is:

$$P_\text{b} = \frac{2\pi\,5.7\,\text{GHz}}{2\pi\,194\,\text{THz}} \cdot \frac{826\,\text{m\,s}^{-1}}{2\pi\,17\,\text{MHz}} \cdot \frac{\pi^2}{(0.024\,\text{m})^2} \cdot \frac{1}{195\,\text{W}^{-1}\text{m}^{-1}} = 20\,\text{nW}. \tag{68}$$

## III. SCATTERING MATRIX FORMULATION

In this section, we present a phenomenological model that captures the behavior of the nonreciprocal modulation produced by nonlocal inter-band Brillouin scattering. This scattering matrix model may also be used to explore the properties of cascaded nonreciprocal circuits.

We begin by representing each of the four ports of the NIBS modulator diagrammed in Fig. 4 as an element of a column vector $\mathbf{A}$. Light at a single port $i$ is represented as:

$$\mathbf{A}_i = \begin{pmatrix} \delta_{1i} \\ \delta_{2i} \\ \delta_{3i} \\ \delta_{4i} \end{pmatrix}. \tag{69}$$

where we have normalized the total power amplitude to a value of 1. We assume idealized mode converters (i.e. neglect cross-talk) and write the scattering matrix that represents the effect of the NIBS modulator on an input signal as:

$$\mathbf{A}_\text{out} = \mathbf{B} \cdot \mathbf{A}_\text{in}, \tag{70}$$

where $\mathbf{A}_\text{in}$ and $\mathbf{A}_\text{out}$ are four-element vectors that represent the respective input and output fields, and $\mathbf{B}$ is defined by

$$\mathbf{B} = \begin{pmatrix} 0 & 0 & \sqrt{1-\eta_\text{f}^2} & \eta_\text{b} e^{\pm i(\phi_\text{b}+\Omega t)} \\ 0 & 0 & \eta_\text{b} e^{\mp i(\phi_\text{b}+\Omega t)} & \sqrt{1-\eta_\text{f}^2} \\ \sqrt{1-\eta_\text{f}^2} & \eta_\text{f} e^{\mp i(\phi_\text{f}+\Omega t)} & 0 & 0 \\ \eta_\text{f} e^{\pm i(\phi_\text{f}+\Omega t)} & \sqrt{1-\eta_\text{f}^2} & 0 & 0 \end{pmatrix}. \tag{71}$$

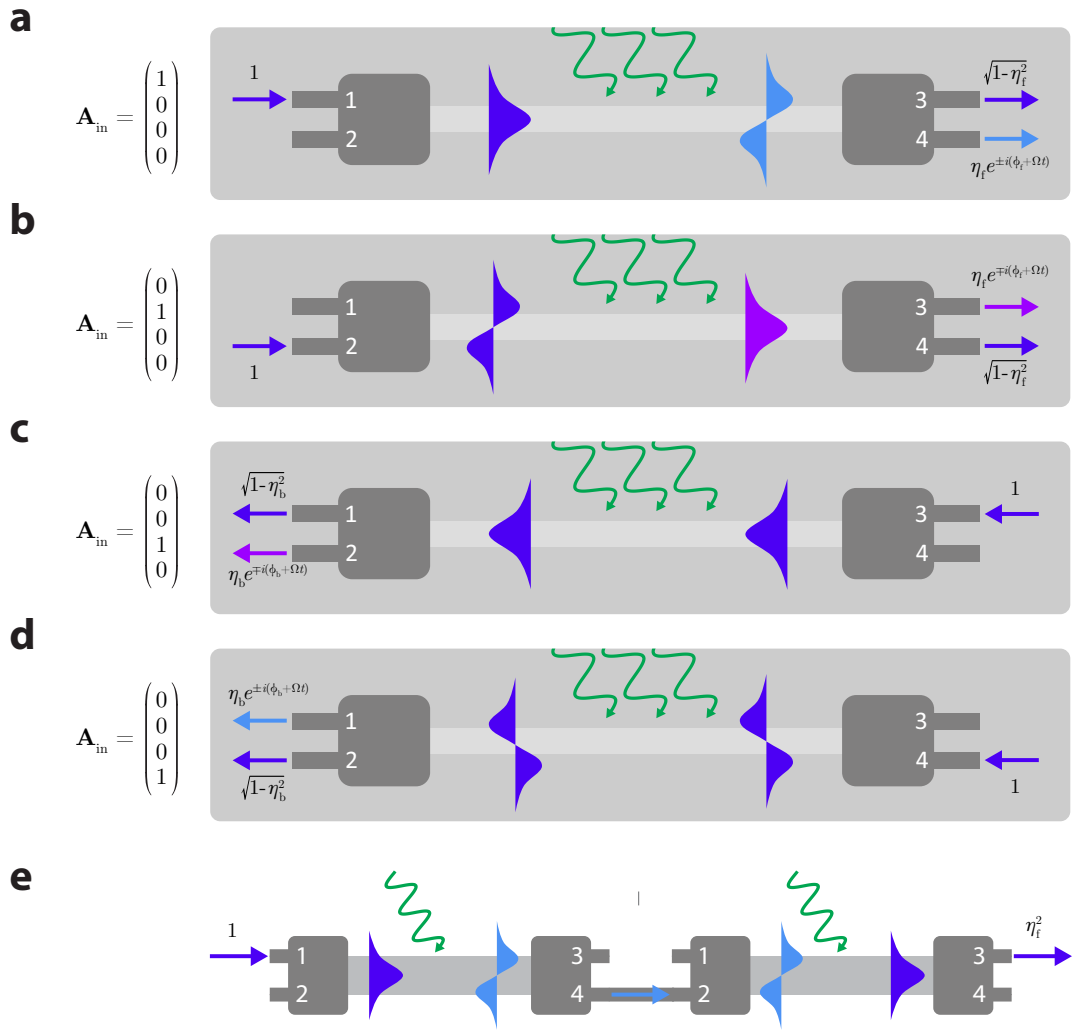

FIG. 4: NIBS modulation within the scattering matrix formulation (a-d) depict input and output wave amplitudes for four different configurations corresponding to injecting light in each port of the four-port NIBS modulator. In (a-b), significant mode conversion is achieved ($\eta_f^2 \approx 1$), while in the backward direction light propagates through the device with only a small amount being mode-converted ($\eta_b^2 \ll 1$). (e) Depicts one approach to create a frequency-neutral isolator by cascading two NIBS modulators.

Here, $\eta_f^2$ and $\eta_b^2$ are the inter-band power conversion efficiencies in the forward and backward directions, respectively. $\phi_f$ and $\phi_b$ are the corresponding phase-shifts associated with the inter-band scattering, and $\Omega$ is the frequency shift imparted by the driven acoustic field. The upper or lower of $\pm$ and $\mp$ represents the case of a forward- or backward-propagating acoustic field, respectively. The antidiagonal terms represent inter-band scattering through Stokes or anti-Stokes processes. When this matrix is asymmetric ($\eta_f \neq \eta_b$), it represents a nonreciprocal mode conversion process.

We assume that $\eta_f^2 \gg \eta_b^2 \approx 0$, i.e. that the device is operating around an optical wavelength $\omega_f$ where strong nonreciprocity is supported in the forward direction. (This same model can be used for strong backward-propagating modulation by considering the case where $\eta_b^2 \gg \eta_f^2 \approx 0$.) This scattering matrix is represented diagrammatically for four different input cases in Fig. 4. The input and output amplitudes at each port are

labeled with their corresponding efficiencies, and in the forward-propagating case strong mode conversion is observed.

In the case of perfect inter-band conversion where $\eta_f^2 \approx 1$, the idealized scattering matrix becomes:

$$\mathbf{B}_{\eta_f^2=1} = \begin{pmatrix} 0 & 0 & 1 & 0 \\ 0 & 0 & 0 & 1 \\ 0 & e^{\mp i(\phi_f + \Omega t)} & 0 & 0 \\ e^{\pm i(\phi_f + \Omega t)} & 0 & 0 & 0 \end{pmatrix}. \tag{72}$$

In this form, the nonreciprocal mode conversion is visible as the antidiagonal terms which are only present in the first two columns. This scattering matrix also represents a frequency-shifting four-port circulator; light incident in port 1 exits through port 4, port 4 maps light to port 2, port 2 maps light to port 3, and port 3 maps light back to port 1. This can be seen through the following scattering matrix equations:

$$\mathbf{B}_{\eta_f^2=1} \cdot \mathbf{A}_1 = e^{\pm i(\phi_f + \Omega t)} \mathbf{A}_4 \tag{73}$$

$$\mathbf{B}_{\eta_f^2=1} \cdot \mathbf{A}_4 = \mathbf{A}_2 \tag{74}$$

$$\mathbf{B}_{\eta_f^2=1} \cdot \mathbf{A}_2 = e^{\mp i(\phi_f + \Omega t)} \mathbf{A}_3 \tag{75}$$

$$\mathbf{B}_{\eta_f^2=1} \cdot \mathbf{A}_3 = \mathbf{A}_1 \tag{76}$$

This scattering matrix formulation can be also be used to consider cascaded arrays of NIBS modulator devices. To consider this case, we introduce an auxiliary matrix $\mathbf{T}_{ij}$ defined by

$$\mathbf{T}_{ij} = \begin{pmatrix} \delta_{1i}\delta_{1j} & \delta_{1i}\delta_{2j} & \delta_{1i}\delta_{3j} & \delta_{1i}\delta_{4j} \\ \delta_{2i}\delta_{1j} & \delta_{2i}\delta_{2j} & \delta_{2i}\delta_{3j} & \delta_{2i}\delta_{4j} \\ \delta_{3i}\delta_{1j} & \delta_{3i}\delta_{2j} & \delta_{3i}\delta_{3j} & \delta_{3i}\delta_{4j} \\ \delta_{4i}\delta_{1j} & \delta_{4i}\delta_{2j} & \delta_{4i}\delta_{3j} & \delta_{4i}\delta_{4j} \end{pmatrix}. \tag{77}$$

In a series of two cascaded devices, $\mathbf{T}_{ij}$ can be used to represent connecting port $i$ of the first device to port $j$ of the second device. A repeated index, i.e. $\mathbf{T}_{ii}$ can be used to represent back-reflecting light at port $i$.

We consider a simple model for a frequency-neutral (non frequency-shifting) isolator consisting of two NIBS modulators diagrammed in Fig. 4e, with port 4 of the first modulator connected to port 2 of the second. In the forward direction, light incident in port 1 is mode-converted through a Stokes process in the first modulator, then converted back to the fundamental mode through an anti-Stokes process in the second modulator. The resulting transmission is:

$$\mathbf{A}_\mathrm{f} = \mathbf{B} \cdot \mathbf{T}_{42} \cdot \mathbf{B} \cdot \mathbf{A}_1$$

$$= \begin{pmatrix} 0 \\ 0 \\ \eta_\mathrm{f}^2 \\ \eta_\mathrm{f}\sqrt{1-\eta_\mathrm{f}^2}e^{\pm i(\phi_\mathrm{f}+\Omega t)} \end{pmatrix}. \tag{78}$$

By contrast, light incident in the backward direction through port 3 of the second modulator does not experience strong mode conversion when $\eta_\mathrm{b} \ll 1$. The resulting transmission in the backward direction is:

$$\mathbf{A}_\mathrm{b} = \mathbf{B} \cdot \mathbf{T}_{24} \cdot \mathbf{B} \cdot \mathbf{A}_3$$

$$= \begin{pmatrix} \eta_\mathrm{b}^2 \\ \eta_\mathrm{b}\sqrt{1-\eta_\mathrm{b}^2}e^{\mp i(\phi_\mathrm{b}+\Omega t)} \\ 0 \\ 0 \end{pmatrix}. \tag{79}$$

The corresponding nonreciprocal power transmission between forward and backward directions is $T_\mathrm{nr} \equiv P_{1\to 3}/P_{3\to 1} = \eta_\mathrm{f}^4/\eta_\mathrm{b}^4$. The effective forward insertion loss is $\eta_\mathrm{f}^4$.

## IV. TUNABILITY OF OPERATION WAVELENGTH

The optically-driven acoustic phonon used to mediate the NIBS process has a wavevector set directly by the difference in wavevectors between the optical drive tones (Section ID). This phonon then modulates light in a separate waveguide at wavelengths where two optical modes exist with the same wavevector difference. In the present work, the pump wavelength was fixed at $\lambda_\mathrm{p} = 1550$ nm to produce modulation over a $\sim 1$ nm bandwidth around the probe wavelength ($\lambda_\mathrm{b}$ or $\lambda_\mathrm{f}$ depending on the direction of injected light).

By changing the pump wavelength, the phonon wavevector (and hence probe wavelength) can be directly tuned. We demonstrate this by adjusting our pump wavelength from $\lambda_\mathrm{p} = 1530$ nm to $\lambda_\mathrm{p} = 1565$ nm to translate the phase-matched modulation wavelength over a similar 35 nm range. As plotted in Fig. 5, as the pump wavelength is tuned, the center modulation wavelength changes by a corresponding amount, with little deviation in the overall shape of the modulation response. This tuning range was limited only by that of the pump laser.

## V. ANTI-STOKES MODULATION DATA

In the present work we have presented mode conversion through Stokes scattering processes. If desired, all of the same physics can be applied to produce single-sideband modulation through an anti-Stokes (blue-shifting) scattering process. Because NIBS mediates mode conversion between a pair of optical modes, as


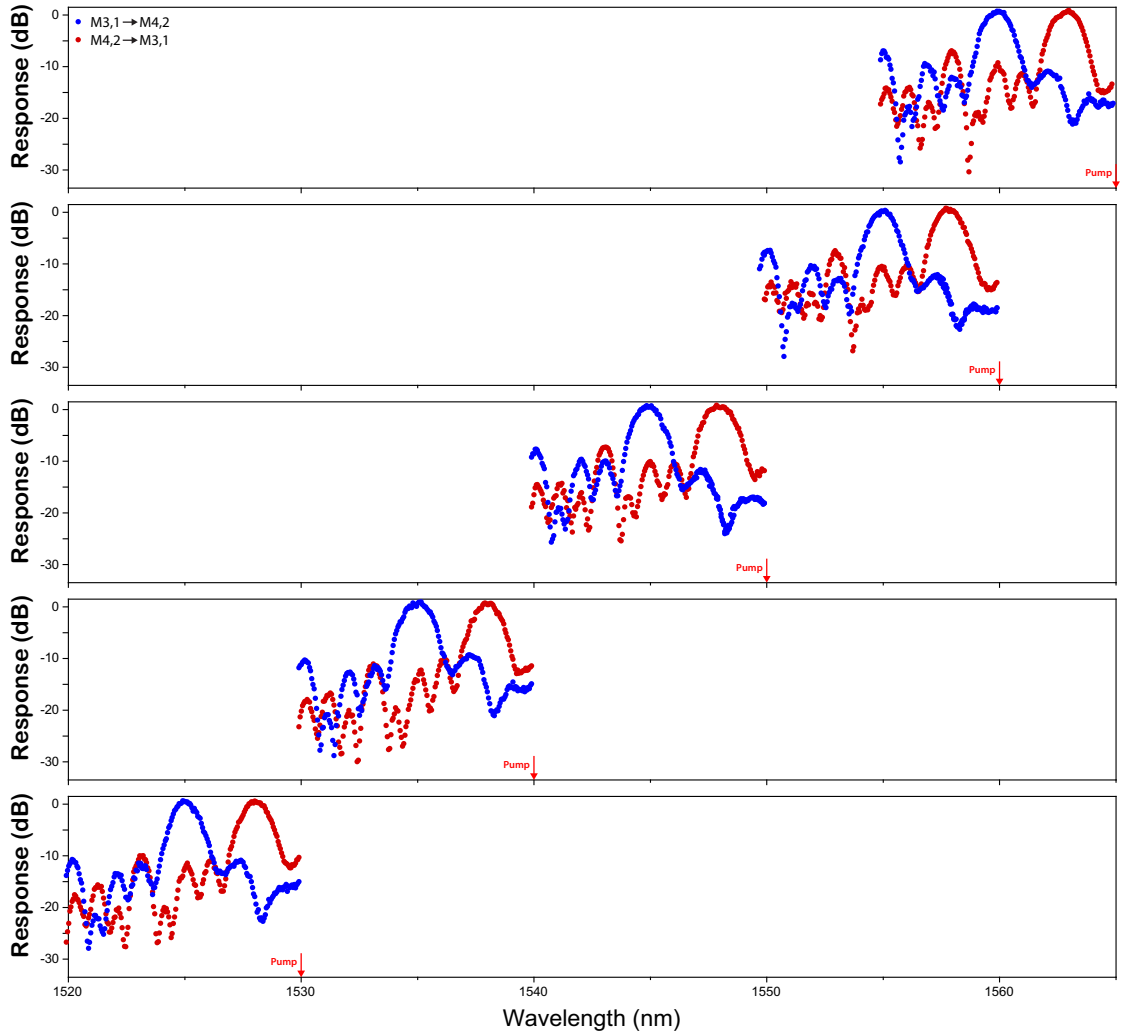

FIG. 5: Tunability of the NIBS modulator wavelength response as the pump (drive waveguide) wavelength is changed. As the pump is tuned from 1565 nm to 1530 nm, the modulator response is shifted down in wavelength by a corresponding amount. Data are plotted for forward (port M3,1→M4,2) operation (blue dots) and backward (port M4,2→M3,1) operation (red dots).

described in Section III, this is achieved by injecting light into the opposite mode as would have been used for a Stokes process, as plotted is plotted in Fig. 6a. The energy level diagram for this process, a form of nonlocal coherent anti-Stokes Brillouin scattering, is plotted in Fig. 6b.

Forward and backward modulation response data for the anti-Stokes process in a NIBS modulator device are plotted in Fig. 6c. These data show the same behavior as for Stokes modulation in the same device. (Fig. 6e). However, the output light is blue-shifted through this process, with the output spectrum plotted in Fig. 6d. In general, the modulation response and wavelength-dependence of phase-matching for the anti-Stokes process are identical to that of the Stokes processes studied in the paper. Hence in situations where this process is desired, the same operation principles may be applied.

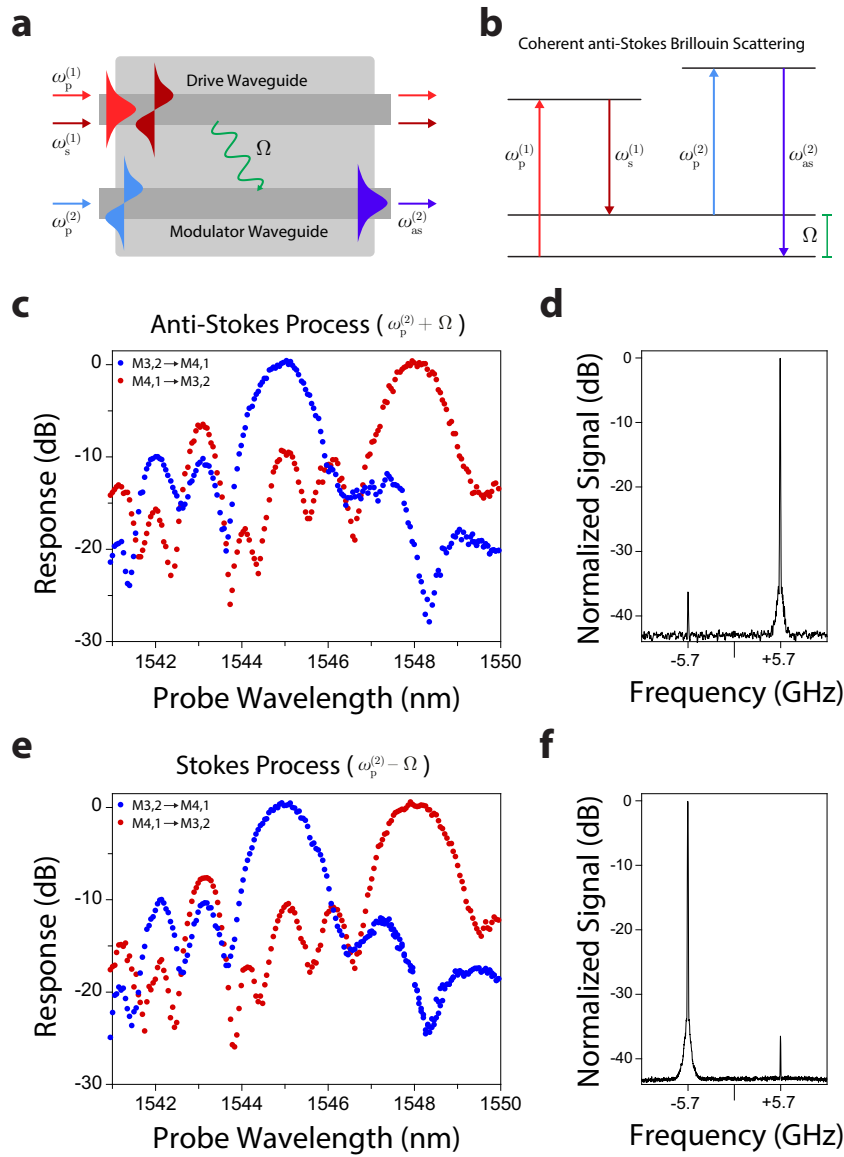

FIG. 6: Anti-Stokes NIBS modulation. (a-b) depict the device operation scheme and energy level diagram in analogy to Fig. 1 of the main article. (a) shows a cartoon of nonlocal inter-band Brillouin scattering for the anti-Stokes process. While the phonon emission is unchanged from the text, light is incident in the anti-symmetric mode of the modulator waveguide. The phonon blue-shifts and mode converts this light as it traverses the device. (b) plots the energy level diagram for this coherent anti-Stokes Brillouin scattering process. (c) shows the anti-Stokes modulation response for light propagating in both directions through the modulator waveguide as a function of probe wavelength. (d) plots the corresponding output spectrum relative to a single optical incident field for a wavelength of 1548.3 nm. (e-f) plot data for the Stokes process in the same device for comparison. (e) plots the forward-backward Stokes modulation response, while (f) plots the output spectrum at a probe wavelength of 1548.3 nm.

## VI. EXPERIMENTAL MODULATION EFFICIENCY

In this section, we explore the dependence of the experimental modulation efficiency $\eta^2 = P_s^{(2)}(L)/P_p^{(2)}(0)$ on the incident drive-waveguide pump powers, $P_p^{(1)}$ and $P_s^{(1)}$, which are guided in the symmetric and anti-symmetric waveguide modes, respectively. Throughout these measurements the relative input ratio of these

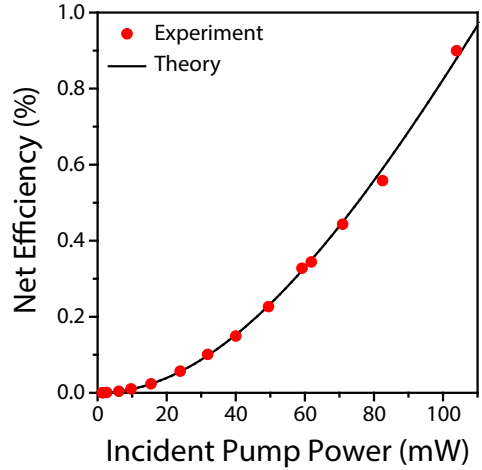

FIG. 7: NIBS modulation efficiency for one device as a function of incident drive-waveguide pump power.

powers is $P_s^{(1)}(0) = 0.65 P_p^{(1)}(0)$, i.e. $k = 0.65$.

Fig. 7 shows the experimentally measured modulation efficiency for one NIBS modulator device as a function of total drive waveguide power up to a maximum of $P = 104$ mW. At the highest tested pump power, the observed modulation efficiency is about 0.9%, including a ∼1 dB reduction in efficiency due to linear loss in the modulator waveguide. The total incident power, and hence modulation efficiency, was limited here by the power available from the EDFAs external to the experiment. A theoretical curve (black) corresponding to a numerical simulation of the full dynamics of the system (Eqs. (23-24,,26-28)) is plotted atop the data.

The numerical curve agrees well with the experimental data with a Brillouin gain coefficient in each waveguide of $G = G^{(1)} = G^{(2)} = 195 \pm 10$ W$^{-1}$m$^{-1}$. The remaining parameters used in these calculations, which are corroborated by independent waveguide measurements, are summarized in Table 1 below. Of these parameters, $L$ is determined through fabrication, two-photon absorption coefficients [17] and acoustic group and phase velocities ($v_{b,g}$ and $v_{b,p}$) are determined through finite element simulations, and all other quantities are determined from experimental measurements.

TABLE I: Measured and Calculated Device Parameters

| Linear optical parameters | value |
|---|---|
| $L$ | 2.387 cm |
| $\alpha_+^{(1,2)}$ | 4.6 m$^{-1}$ |
| $\alpha_-^{(1,2)}$ | 9.2 m$^{-1}$ |
| **Nonlinear optical parameters** | **value** |
| Brillouin Gain | |
| $G$ | 195 W$^{-1}$m$^{-1}$ |
| TPA coefficients | |
| $\beta_{++}^{(1,2)}$ | $34 \pm 10$ m$^{-1}$W$^{-1}$ |
| $\beta_{--}^{(1,2)}$ | $30 \pm 9$ m$^{-1}$W$^{-1}$ |
| $\beta_{+-}^{(1,2)} = \beta^{21}$ | $20 \pm 6$ m$^{-1}$W$^{-1}$ |
| FCA coefficients | |
| $\gamma_{+++}^{(1,2)}$ | $1000 \pm 400$ m$^{-1}$W$^{-2}$ |
| $\gamma_{---}^{(1,2)}$ | $790 \pm 430$ m$^{-1}$W$^{-2}$ |
| $\gamma_{+--}^{(1,2)} \approx \gamma_{-++}^{122}$ | $340 \pm 200$ m$^{-1}$W$^{-2}$ |
| **Acoustic parameters** | **value** |
| $\Omega_B$ | $2\pi$ 5.70 GHz |
| $\Gamma$ | $2\pi$ 17.0 MHz |
| $Q = \Omega_B/\Gamma$ | 335 |
| $q$ | $4.5 \times 10^5$ m$^{-1}$ |
| $v_{b,g}$ | 826 m/s |
| $v_{b,p}$ | $8.4 \times 10^4$ m/s |

[1] Yu, Z. & Fan, S. Complete optical isolation created by indirect interband photonic transitions. *Nature Photon.* **3**, 91–94 (2009).

[2] Huang, X. & Fan, S. Complete all-optical silica fiber isolator via stimulated Brillouin scattering. *J. Lightwave Technol.* **29**, 2267–2275 (2011).

[3] Kang, M.S., Butsch, A. & Russell, P.St.J. Reconfigurable light-driven opto-acoustic isolators in photonic crystal fibre. *Nat. Photon.* **5**, 549–553 (2011).

[4] Kittlaus, E.A., Otterstrom, N.T. & Rakich, P.T. On-chip inter-modal Brillouin scattering. *Nat. Commun.* **8**, 15819 (2017).

[5] Otterstrom, N.T., Behunin, R.O., Kittlaus, E.A., Wang, Z. & Rakich, P.T. A silicon Brillouin laser. *arXiv preprint arXiv:1705.05813* (2017).

[6] Wolff, C., Van Laer, R., Steel, M.J., Eggleton, B.J. & Poulton, C.G. Brillouin resonance broadening due to structural variations in nanoscale waveguides. *New J. Phys.* **18**, 025006 (2016).

[7] Yin, L., Lin, Q. & Agrawal, G.P. Dispersion tailoring and soliton propagation in silicon waveguides. *Opt. Lett.* **31**,

1295–1297 (2006).

[8] Liu, X. *et al.* Conformal dielectric overlayers for engineering dispersion and effective nonlinearity of silicon nanophotonic wires. *Opt. Lett.* **33**, 2889–2891 (2008).

[9] Zhang, L., Yue, Y., Beausoleil, R.G. & Willner, A.E. Flattened dispersion in silicon slot waveguides. *Opt. Express* **18**, 20529–20534 (2010).

[10] Poulton, C.G. *et al.* Design for broadband on-chip isolator using stimulated Brillouin scattering in dispersion-engineered chalcogenide waveguides. *Opt. Express* **20**, 21235–21246 (2012).

[11] Zhang, L. *et al.* Silicon waveguide with four zero-dispersion wavelengths and its application in on-chip octave-spanning supercontinuum generation. *Opt. Express* **20**, 1685–1690 (2012).

[12] Liang, H., He, Y., Luo, R. & Lin, Q. Ultra-broadband dispersion engineering of nanophotonic waveguides. *Opt. Express* **24**, 29444–29451 (2016).

[13] Kharel, P., Behunin, R.O., Renninger, W.H. & Rakich, P.T. Noise and dynamics in forward brillouin interactions. *Phys. Rev. A* **93**, 063806 (2016).

[14] Wolff, C., Gutsche, P., Steel, M.J., Eggleton, B.J. & Poulton, C.G. Impact of nonlinear loss on stimulated Brillouin scattering. *J. Opt. Soc. Am. B* **32**, 1968–1978 (2015).

[15] Tadesse, S.A. & Li, M. Sub-optical wavelength acoustic wave modulation of integrated photonic resonators at microwave frequencies. *Nat. Commun.* **5**, 5402 (2014).

[16] Sohn, D.B., Kim, S. & Bahl, G. Time-reversal symmetry breaking with acoustic pumping of nanophotonic circuits. *Nat. Photonics* **12**, 91 (2018).

[17] Afshar V., S. & Monro, T.M. A full vectorial model for pulse propagation in emerging waveguides with subwavelength structures part I: Kerr nonlinearity. *Opt. Express* **17**, 2298–2318 (2009).



\end{document}